\documentclass[aps,letterpaper,twocolumn,preprintnumbers,floatfix,superscriptaddress,nofootinbib]{revtex4-1}

\usepackage{amsmath}
\usepackage{amsfonts}
\usepackage{amssymb}
\usepackage{graphicx, rotating}
\usepackage{epstopdf}
\usepackage{epsfig}
\usepackage{latexsym}
\usepackage{color}
\usepackage[dvipsnames]{xcolor}
\usepackage{multirow}

\usepackage{xcolor,colortbl}

\usepackage{tikz-feynman}
\tikzfeynmanset{warn luatex=false}

\usepackage[skip=0.333\baselineskip]{caption}

\usepackage{slashed}
\usepackage{hyperref}
\hypersetup{colorlinks, citecolor=bluscuro, linkcolor=bluscuro, urlcolor=bluscuro}
\definecolor{rossos}{cmyk}{0,1,1,0.55}
\definecolor{bluscuro}{rgb}{0.15, 0.2, .85}
\definecolor{bluchiaro}{cmyk}{1,.3,0.,0.1}
\definecolor{verdescuro}{rgb}{0.3,0.8,0.3}

\newcommand{\bs}{\hat s}

\newcommand{\os}{{\hat s^\prime}}

\newcommand{\doa}{\hspace{-3pt}\footnotesize\begin{array}{c}\smallfrown\\[-7pt]\smallsmile\end{array}}

\newcommand{\eq}[1]{Eq.~(\ref{#1})}

\newcommand{\nn}{\nonumber}

\newcommand{\be}{\begin{equation}}
\newcommand{\ee}{\end{equation}}          
\newcommand{\bea}{\begin{eqnarray}}
\newcommand{\eea}{\end{eqnarray}}
\newcommand{\bc}{\begin{center}}
	\newcommand{\ec}{\end{center}}

\newcommand{\M}{{\cal M}}
\newcommand{\Mh}{{\hat{\cal M}}}

\begin{document}

\widetext

\begin{flushright}
{\small 
Saclay-t21/094}
\end{flushright}

\title{The IR-Side of Positivity Bounds}

\author{Brando Bellazzini}
\affiliation{Universit\'e Paris-Saclay, CNRS, CEA, Institut de Physique Th\'eorique, 91191, Gif-sur-Yvette, France. }
\affiliation{Theoretical Particle Physics Laboratory (LPTP), Institute of Physics, EPFL, Lausanne, Switzerland}
\affiliation{CERN, Theoretical Physics Department, Geneva, Switzerland}
\author{Marc Riembau}
\affiliation{Theoretical Particle Physics Laboratory (LPTP), Institute of Physics, EPFL, Lausanne, Switzerland}
\author{Francesco Riva}
\affiliation{D\'epartment de Physique Th\'eorique, Universit\'e de Gen\`eve,
24 quai Ernest-Ansermet, 1211 Gen\`eve 4, Switzerland}

\begin{abstract}
\noindent 
We show how calculable IR loop effects  impact positivity bounds in Effective Field Theories with causal and unitary UV completions.
We identify  infrared singularities which appear in dispersion relations at $|t|\lesssim m^2$. In the massless limit, they weaken two-sided bounds based on crossing symmetry, such as the lower bound on the amplitude for Galileon scattering.
For amplitudes that are analytic in $s$ even for large negative $t$, i.e. $|t|\gg m^2$, we propose a new simple analytic approach to dispersive bounds, which are instead insensitive to the singularities, and explicitly compute the finite contributions from loops. Finally we show that the singularity do not affect the bounds based on smearing in impact parameter.

\end{abstract}

\maketitle
\medskip


\section{Introduction}

The space of low energy relativistic effective field theories (EFT) is subject to fundamental constraints known as positivity bounds.  They encode, in the form of inequalities among scattering amplitudes  evaluated in the infrared (IR),  the conditions that EFTs emerge from the renormalization-group~(RG) evolution of  unitary and causal microscopic dynamics. 
 IR-consistent EFT failing the positivity bounds belong   to the ``swampland''.

Positivity bounds stem from dispersion relations that provide this  UV/IR connection~\cite{Pham:1985cr,Ananthanarayan:1994hf,Pennington:1994kc}. While rooted in the old S-matrix bootstrap, the deeper understanding and implications  of positivity bounds was put forward in \cite{Adams:2006sv}, where the connection with the requirement of subluminal propagation of the low-energy degrees of freedom was also made explicit. 
 
 Generalization to arbitrary spinning particles \cite{Bellazzini:2016xrt,deRham:2017zjm} and to finite scattering angle \cite{Vecchi:2007na,Nicolis:2009qm,Bellazzini:2014waa,deRham:2017avq}, along with interesting applications in particle phenomenology \cite{Distler:2006if,Manohar:2008tc,Low:2009di,Falkowski:2012vh,Bellazzini:2017bkb,Bellazzini:2014waa,Bellazzini:2018paj,Zhang:2018shp,Remmen:2019cyz,Bellazzini:2019bzh,Remmen:2019cyz,Englert:2019zmt,Trott:2020ebl,Bonnefoy:2020yee,Davighi:2021osh,Chala:2021wpj,Azatov:2021ygj},  and in gravitational physics \cite{Gruzinov:2006ie,Bellazzini:2015cra,Bellazzini:2017fep,Hamada:2018dde,deRham:2018qqo,Alberte:2019xfh,Bellazzini:2019xts,Kim:2019wjo,Tokuda:2020mlf,Herrero-Valea:2020wxz,Bern:2021ppb}, or in connection to the positivity of time delay \cite{Camanho:2014apa,Afkhami-Jeddi:2018apj,Bellazzini:2021shn},   have recently lead to the uncovering of the  architecture behind all positivity bounds, (called EFT-Hedron in \cite{Arkani-Hamed:2020blm}),  where infinitely many Wilson coefficient are constrained, at tree-level, by double-sided bounds~\cite{Bellazzini:2020cot,Tolley:2020gtv,Caron-Huot:2020cmc,Arkani-Hamed:2020blm,Sinha:2020win,Caron-Huot:2021rmr,Chiang:2021ziz,Henriksson:2021ymi}.  
 
 The chief purpose of this article is to show how calculable IR effects and loops deform the EFT-Hedron beyond the idealized tree-level limit. 
 Discussing realistic EFTs at finite coupling in the IR is important in order to render positivity bounds actually sharp: throwing theories in the swampland  must be irreversible, should not be undone by  weak coupling corrections possibly enhanced by IR divergences.

Broadly speaking, there are two implications of IR loop effects  on positivity bounds for Wilson coefficients, when evaluated at the EFT cutoff where 2-sided bounds are most relevant. 
First of all, $n$-subtracted dispersion relations involve (via the  integral along the IR discontinuity) all EFT couplings, in contrast to the tree-level limit where they select  the coefficient of $s^n$ in the amplitude.
Therefore sharp dispersive bounds deliver not so sharp  constraints on Wilson coefficients, and for practical applications require perturbative assumptions about the EFT convergence~\cite{Bellazzini:2020cot}. 
Secondly, and perhaps more importantly,  in the massless limit the non analyticities extend to  $s\to 0$ and $t\to 0$. There, the most relevant interactions and their IR loops always dominate the amplitude, screening, potentially, information on less relevant couplings.   
Moreover, the amplitude being non-analytic in $t$ as the mass goes to zero restricts the use of the dispersion relations based on the near-forward region. 

In this article  we present a thorough study of 1-loop effects within the EFT of a scalar with 4-point interactions, focusing on forward singularities and their impact on 2-sided bounds. 
As a case study we discuss bounds on the ratio $g_{3,1}s/g_{2}$ between the coefficients $g_{3,1}$ of $stu$ in the amplitude (which would-be dominant for Galileons  \cite{Nicolis:2008in}) and the coefficient $g_2$ of $s^2+t^2+u^2$ (ordinary Goldstone boson), and discuss how the arguments generalize to other couplings. 

Expressing \emph{near-forward} $t\approx0$ dispersion relations as 2D-moments of a positive measure in the UV, generalizing \cite{Bellazzini:2020cot} along the lines of \cite{Chiang:2021ziz}, we show that the  lower bound on $g_{3,1}s/g_{2}$ that holds at tree level actually disappears, in the massless limit, regardless of how small the coupling is taken,  as shown in Fig.~\ref{figm}. 
The same conclusion holds for the approaches of Refs.~\cite{Arkani-Hamed:2020blm,Tolley:2020gtv,Caron-Huot:2020cmc}. 

This pessimistic result is however based on the near-forward positivity bounds only, that posit amplitude's analyticity only within the  rigorously proven domain. We contrast such conclusions with those we obtain by working with an enlarged domain of analyticity in $s$ for  negative $t\ll -m^2$ which, despite not being rigorously proved, is supported by the perturbative analysis of Landau equations within the EFT. We present a simple  approach to dispersion relations that shows that the tree-level bounds do actually survive also at loop level, as long as the coupling remains perturbative, under the assumption of maximal analyticity.  
Our approach is analytic and simple,  and nicely complements the numerical technique of \cite{Caron-Huot:2021rmr}.

\section{Dispersion Relations at Finite $t$}
We study  $2\to2$ scattering of a single particle of spin-0. We  assume that the amplitude $\M(s,t)$  is analytic  in the entire complex $s$ plane except the physical branch-cut at $s\geq 4m^2$ and its crossing symmetric counterpart at $s<-t$, for fixed values of~$t$ as specified in the following sections.
We focus on the  amplitude with $n+2$ ($n\geq0$) subtractions in $s=0$ and $n+1$ subtractions in $s=-t$, and define ``arcs'' the following $\bs$- and $t$-dependent contour integral
\begin{equation}\label{archdeft}
a_n(\bs,t)\equiv \int_{\doa}\frac{d\os}{2\pi i\os}\frac{\Mh(\os,t)}{[\os(\os+t)]^{n+1}} \,,\quad n\geq0
\end{equation}
where  
\begin{equation}
\bs\equiv s-2m^2\,,\qquad \Mh(\hat s,t)\equiv \M(s,t)
\end{equation}
and ${\doa}$ is the circle with radius $\bs +t/2$ and centered at~$-t/2$ (minus its interception with the real axis). 
Arcs probe the theory at energy $\bs$ and momentum transfer $q^2=-t$, and are suited to capture the RG flow within~EFTs.

We deform the contour $\doa$ into a contour that encompasses the discontinuities on positive $\os\geq \bs$ and negative $\os\leq -\bs-t $ real axis, together with upper and lower semicircles at infinity. 
We further assume that the latter vanish, because, 
\begin{equation}
\lim_{|s|\to \infty} \M(s,t)/s^2 =0 
\end{equation}
as implied by the Froissart-Martin bound \cite{Froissart:1961ux,Martin:1962rt,Jin:1964zza}. 
From  crossing symmetry, real analyticity, and the partial-wave expansion,
 \begin{gather}
 \Mh(\bs,t)=  \Mh(-\bs-t,t)\,, \quad
 \Mh(\bs,t)=  \Mh^*(\bs^*,t)\,,\nn\\ 
 \Mh(\bs,t)= \sum_{l=0}^\infty P_\ell(1+\frac{2t}{\bs\!\!-\!2m^2})\hat f_\ell(\bs)
 \end{gather}
where $P_\ell(\cos\theta)$ are the Legendre Polynomials and $\ell$ is even for identical scalars,  we can express the \emph{arcs} in terms of their UV integral representation,
\begin{eqnarray}
a_n(\bs,t)
&=&\frac{2}{\pi}\int_{\bs}^\infty \sum_{\ell=0}^\infty  d\os\textrm{Im} \hat f_\ell(\os)  \frac{I^n_t(\ell,\os)}{\hat s^{\prime2n+3}}
\label{archdeftUV}
\end{eqnarray}
with the kernel $I^n_t(\ell,\os)$ given by,
\begin{equation}\label{integrandI}
I^n_t(\ell,\os)\equiv  \frac{\left(1+\frac{t}{2\os}\right)}{\left(1+\frac{t}{\os}\right)^{n+2}} P_\ell(1+\frac{2t}{\os-2m^2})\,.
\end{equation}
In the following we are interested in the limit 
\begin{equation}
m^2\ll s
\end{equation}
where also $\bs\to s$.  In practice we will set $m\rightarrow 0$ everywhere except in the presence of IR divergences where the mass acts as an explicit regulator.

\subsection{IR Arcs}

Arcs  can be computed  within the EFT via \eq{archdeft} in terms of  Wilson coefficients. 
Consider and EFT which, in a weak-coupling regime,  can be well approximated by the tree-level expression,
\begin{equation}\label{amptreeXY}
\M(s,t)=\sum_{n,q} g_{n,q} \left(\frac{s^2 + t^2 + u^2}{2}\right)^{\frac{n-3q}{2}}\cdot \left(s  t u\right)^{q}
\end{equation}
where $n$ sets the overall energy-squared scaling, whereas $q$ tells how fast the amplitude vanishes for $t\rightarrow 0$ and fixed $s$. 
Then, at $t=0$ the arcs read 
\begin{equation}
a_n(\bs,0)=g_{2n+2,0}
\end{equation}
 while at finite~$t$,\footnote{\label{footnote1}The subtraction choice $\frac{1}{[s(s+t)]^{n+1}}=\frac{(-t)^{n+1}}{(stu)^{n+1}}$ in \eq{archdeft} implies that the coefficient of $(stu)^q$ only appears in $a_{n\geq q-1}$, while $(s^2+t^2+u^2)^p$ appears in all $a_{n\leq p}$. This coincides with the choice of Ref.~\cite{Bellazzini:2020cot} only at $t=0$.}
\begin{equation}\label{arcsexplicit}
a_0(\bs,t)=\sum_{n=1}^\infty[ n t^{2n-2}g_{2n,0}- t^{2n-1}g_{2n+1,1}]\,.
\end{equation}
with similar expressions for higher arcs. At tree-level, arcs are in one-to-one correspondence with couplings. In particular, knowledge of
\begin{equation}\label{minimalarcs}
a_n|_{t=0},\partial_t a_n|_{t=0},\cdots,\partial_t^{n+1} a_n|_{t=0}
\end{equation}
is enough to reconstruct the whole series of Wilson coefficients.

\vspace{2mm}
From \eq{amptreeXY}, IR loop effects are  calculable. 
In this article we will focus on the most important 1-loop effect, from two insertions of $g_{2,0}$:
\begin{equation}\label{amploop}
\delta\M=\frac{\beta_4}{2}s^2(s^2-\frac{tu}{21})\log(-s)+(s\leftrightarrow t,u)\,,
\end{equation}
where the renormalisation scale is implicit and,
\begin{equation}
\beta_4=-\frac{7}{10}\frac{g_{2,0}^2}{16\pi^2}
\end{equation}
which must be small $|\beta_4 \bs^2|\ll 1$ for the perturbative expansion to make sense.
This gives an additional contribution to the first arc,
\begin{eqnarray}\label{deltaa0}
\delta a_0(\bs,t)&=&\beta_4\big(\bs^2 - \frac{41}{21}\bs t + \frac{t^2}{2} - \frac{\bs^3}{2 (\bs + t)}\\
&&+t^2  (2 \log (\bs+t) + \frac{1}{42}\log \frac{-t}{\bs+t} )\big)\,,\nn
\end{eqnarray}
which is non-analytic in either $s,t,u\to0$. In particular, the term proportional to $t^2\log -t$ will play an important role in what follows, since present  bounds rely on arc $t$-derivatives in the forward limit, $\partial_t^ma_n|_{t=0}$.
In fact all couplings generate non-analyticities in $t=0$.
The most singular effects in the 1-loop contributions  proportional to the marginal coupling $g_{0,0}$ are,
\bea
\delta\M=&&
\frac{g_{0,0}}{32\pi^2}\,\log\frac{-t}{s}\Big(-g_{0,0}\\
&&+t^2\frac{5g_{2,0}}{3}
+t^3 \frac{ g_{3,1}}{3}
+t^4 \frac{7g_{4,0}}{5}+\cdots \Big)\,.\nn
\eea
These (and all other  1-loop contributions involving~$g_{0,0}$) carry no powers of $s$, and therefore do not appear in arcs and do not enter in the dispersion relations.

On the other hand, 1-loop effects involving less relevant couplings have  singularities:
\begin{align}
\delta\mathcal{M}= -\frac{s^2t^2}{16\pi^2}\log\frac{-t}{s}\Big(-t\frac{g_{2,0}g_{3,1}}{30}+t^2\frac{2g_{2,0}g_{4,0}}{35}\nn\\+t^2\frac{g_{3,1}g_{3,1}}{60}-t^3\frac{2g_{3,1}g_{4,0}}{35}+t^2s^2\frac{g_{4,0}^2}{1260} +\cdots\Big)\, .\nn
\end{align}
Of  these, the first four will appear in the first arc $a_0(s,t)$ and will be subdominant to the term $\propto g_{2,0}^2$ from \eq{deltaa0}. Only the term $\propto g_{4,0}^2$ has enough powers of $s$ to appear in the second arc $a_1$, but it is proportional to $t^4\log(-t)$. Therefore, the second arc and its first three derivatives are regular at $t\to 0$. This trend propagates to higher arcs, with leading divergences,
\begin{equation}\label{singha}
\delta a_n(\bs,t)\propto t^{2n+2}\log\frac{-t}{\bs}\,.
\end{equation}
We deduce that arc $n$ and its first $2n+1$ derivatives are regular in $t\to 0$. This is important in light of \eq{minimalarcs}: it is possible to reconstruct all coefficients in the forward limit, without encountering $t=0$ singularities.\footnote{Note that the pattern of IR divergences \eq{singha} translates into a pattern of UV divergences, as $\partial_t^{(k)}a_n(\bs,t)|_{t=0}$ is mapped to a UV integral proportional to $\sum_\ell \ell^{2k}/s^{n}$. At fixed $n$, for high enough $k$, the IR part diverges, implying a measure with support on arbitrarily large $\ell$. At fixed $k$, for high enough $n$, the IR  converges, implying that contributions from  large $\ell$ also have large $s$. With the ansatz $m^2\sim \ell^\alpha$ for the UV spectrum,  \eq{singha} fixes $\alpha=2$ exactly ($\alpha=1$ would correspond to $a_n(\bs,t)\propto t^{n+1}\log\frac{-t}{\bs}$).} We illustrate this in Table~\ref{tab12}.

\begin{table}
\includegraphics[width=0.45\textwidth]{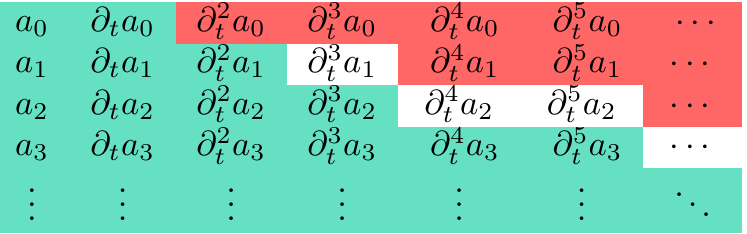}
\caption{{\it In green, a schematic representation of the arcs and their derivatives (both at $t=0$) needed to reconstruct all Wilson coefficients, \eq{minimalarcs}. Red cells  correspond  instead to arc derivatives that are singular at $t=0$, according to \eq{singha}. } }\label{tab12}
\end{table}

\section{2D Moments and Dispersion relations around $t=0$}

We now discuss bounds on arcs, and how they translate to bounds on Wilson coefficients, given the results of the previous section.
In this first section we take a conservative approach and rely  only on analyticity of the amplitude for $t$ within the domain established (assuming unitarity) by Martin \cite{Martin:1965jj,Martin:1969ina,Sommer:1970mr}, whose size is set by the scattered particle mass $m$ . In the  limit of small mass $m\to0$, which we consider here, analyticity requires~$t\to0$. 

One efficient way of deriving near-forward bounds,  involves mapping UV arcs to moments of a positive distribution in $\bs$ and $\ell$~\cite{Bellazzini:2020cot,Chiang:2021ziz}.
Bounds on moments then provide  bounds from the UV,
see~\cite{Chiang:2021ziz,lasserreJB}.
The interpretation in terms of moments is possible because the coefficients in the $t$-Taylor expansion of the arc integrand \eq{integrandI}, are themselves polynomials in   $1/\os$ and $J^2\equiv\ell(\ell+1)$,
\begin{equation}
I^n_{t}(\ell,\os)=\sum_{m=0}^\infty \left(\frac{t}{\os}\right)^m \sum_{k=0}^m d_{n,m-k} \sum_{i=0}^k \alpha^k_i J^{2i}
\end{equation}
with
\bea
d_{n,i}&\equiv&(-1)^i \frac{ (i+n)!}{i! n!}\frac{2n+i+2}{2n+2}\,,\\
\sum_{i=0}^k \alpha^k_i J^{2i}&\equiv& \frac{1}{(k!)^2}\prod_{j=0}^{k-1}(J^2-j(j+1))
\eea
from expanding respectively the ratio and Legendre polynomials in \eq{integrandI}.
Changing variables,
\begin{equation}
x=(\bs/\os)\,,
\end{equation}
we can thus write the arc $t$-derivatives  as linear combinations,
\begin{equation}\label{arcsdersmoms}
\frac{1}{m!}\partial_t^m a_n(t,s)\big|_{t=0} = \sum_{k=0}^m d_{n,m-k} \sum_{j=0}^k \alpha^k_j \,\,\mu^j_{2n+m}
\end{equation}
of two-dimensional moments,
\begin{equation}\label{2dmoments}
\mu_n^q=\frac{1}{\bs^{n+2}} \int d\mu(x,J^2) \,\,\,x^{n} J^{2q} 
\end{equation}
with respect to a positive measure $d\mu(x,J^2)$ (proportional to $x \textrm{Im}\hat f_{\ell(J^2)}(\bs/x) dx >0$ by unitarity) with support on $x\in[0,1]$, times the non-compact discrete set of positive integer numbers (for even $\ell$, $J^2=0,6,20,\cdots$).  
More explicitly, the arcs $a_n(\bs,t)$ at finite $t$ are written via their UV representation as linear combinations of 2D moments, 
\begin{eqnarray}\label{arcsasmoments}
&&a_n(\bs,t)=\mu_{2n}^0+t\left(\mu_{2n+1}^1-\frac{3+n}{2}\mu_{2n+1}^0\right)\\
&&+t^2\left(  \frac{\mu_{2n+2}^2}{4}  -(2+n)\mu_{2n+2}^1 +\frac{(2+n)^2}{2}\mu_{2n+2}^0\right)+\cdots \nn
\end{eqnarray}
where dots denote higher powers of $t$.

\subsection{Bounds on moments} 
Once  arcs are expressed in terms of moments, their positivity constraints stem from the bounds on the 2D moments, which, in turn, are in one-to-one correspondence with the space of all  polynomials in $x$ and $J^2$ that are positive on the integration domain. Indeed, every positive polynomial $\sum \bar \beta_{i,j} x^iJ^{2j}\equiv p(x,J^2)>0$ implies a positivity condition among moments,  $\sum  \beta_{i,j} \mu_{i}^{j}=\int d\mu(x,J^2) p(x,J^2)>0$.

We will first take the limit in which $J^2$ is continuous, which allows us to find \emph{conservative} bounds in terms of a \emph{finite} number of conditions. The continuum bounds (which are quantitatively very similar to exact bounds) will be more conservative because the space of positive polynomials in the continuum contains those positive in the discrete. Moreover, this approach will enable us to obtain analytic bounds without having to rely on numerical extrapolations to large~$\ell$. In appendix \ref{app:bf}, we will show how to include the countably infinite conditions that define the bounds for $\ell$ discrete.

The set $(x,J^2)\in [0,1]\times \mathbb{R}^+$ can be described by the conditions,
\begin{equation}\label{boundariesSchmudigen}
x\geq0\,,\quad 1-x\geq0\,,\quad x J^2\geq 0\,.
\end{equation}
From this, a theorem due to Schm\"udgen~\cite{schmudgen91,lasserreJB} classifies all positive polynomials $p(x,J^2)$ in a domain in  terms of squares of polynomials as,\footnote{Eqs.~(\ref{boundariesSchmudigen},\ref{pospol}) parametrize polynomials of $(x,xJ^2)$ rather than $(x,J^2)$. This is an efficient set of polynomials to characterize moments $\mu_n^q$ with $n\geq q$, as they appear in arcs \eq{arcsdersmoms}. It  will provide simpler expression  when   considering truncations to polynomials of finite order (a finite number of moments), but does not make any difference once polynomials of arbitrary order are taken into account.}
\begin{equation}\label{pospol}
p(x,xJ^2)=\sum_{k} q_k(x,J^2)\left(\sum_{i,j} \beta^k_{i,j} x^{i+j}J^{2j}\right)^2 
\end{equation}
where $q_k$ stem from products of the  monomials  in \eq{boundariesSchmudigen} defining the domain as $q_k\geq0$, and in our case belongs to the set,
\begin{equation}
\{1,x,xJ^2,J^2x^2,1-x,x(1-x),J^2x(1-x),J^2x^2(1-x)\}\,.\nn
\end{equation}
 These capture terms that cannot be written as squares (recall $J^2$ rather than $J$ is the variable entering in the  polynomials). The conditions for each individual $q_k$ can be written in compact form. 
For instance, for $q_k=1$, integrating \eq{pospol} against the positive measure leads to the condition
$\sum_{i,j,m,n} \beta^1_{i,j}\mu^{j+n}_{i+j+m+n} \beta^1_{m,n}>0$, which implies positive definiteness of the infinitely sized matrix,

\begin{equation}\label{hankellocalised}
H_{(0,0)}
=
\left(\begin{array}{c|cc|ccc|c}
\mu_{0}^{0}& \mu_{1}^{0} & \mu_{1}^{1} & \mu_{2}^{0} & \mu_{2}^{1} & \mu_{2}^{2}&\cdots \\\hline
\mu_{1}^{0}  &\mu_{2}^{0} & \mu_{2}^{1} & \mu_{3}^{0} & \mu_{3}^{1} & \mu_{3}^{2}&\cdots \\
\mu_{1}^{1} & \mu_{2}^{1} & \mu_{2}^{2} & \mu_{3}^{1} & \mu_{3}^{2} & \mu_{3}^{3}  &\cdots\\\hline
\mu_{2}^{0} & \mu_{3}^{0} & \mu_{3}^{1} & \mu_{4}^{0} & \mu_{4}^{1} & \mu_{4}^{2}  &\cdots\\
\mu_{2}^{1} & \mu_{3}^{1} & \mu_{3}^{2} & \mu_{4}^{1} & \mu_{4}^{2} & \mu_{4}^{3}  &\cdots\\
 \mu_{2}^{2} & \mu_{3}^{2} & \mu_{3}^{3} & \mu_{4}^{2} & \mu_{4}^{3} & \mu_{4}^{4}&\cdots\\\hline
 \cdots&\cdots&\cdots&\cdots&\cdots&\cdots&\cdots
\end{array}\right)\succ 0
\end{equation}
where the indices of $H_{(0,0)}$ denote the indices of the first entry, and the blocks correspond to monomials of given order $k$ in $x$: $x^{k-i}(xJ^2)^i, i=0,\dots,k$.

Taking into account the other $q_k$, we find the conditions on the shifted matrices,
\begin{gather}
H_{(0,0)}\succ 0\,\quad H_{(1,0)}\succ 0\,\quad H_{(1,1)}\succ 0 \quad H_{(2,1)}\succ 0 \nn\\
H_{(0,0)}-\bs^2 H_{(1,0)}\succ 0 \,,\quad
H_{(1,0)}-\bs^2H_{(2,0)}\succ 0 \,,\quad\nn\\
H_{(1,1)}-\bs^2H_{(2,1)}\succ 0 \,.\label{condsmompos}
\end{gather}

Contrary to the 1D moment problem \cite{Bellazzini:2021shn}, in 2D and higher there is no optimal solution involving only a finite number of moments (the truncated 2D moment problem is solved only asymptotically). This means that in order to find the exact bounds satisfied by, e.g., moments up to order two in $x$ and $xJ^2$, $\{\mu_{0}^{0},\mu_{1}^{0},\mu_{1}^{1},\mu_{2}^{0},\mu_{2}^{1},\mu_{2}^{2}\}$, we still need to compute infinitely many bounds involving infinitely many moments, and then project into the finite subset. Because of Sylvester's criterion, positive definiteness in \eq{condsmompos} implies also positive definiteness of all finite-size principal minors, corresponding to matrices $H_{(i,j)}$ of finite size.  This necessary but not sufficient condition leads to conservative bounds.
For instance, the bounds in \eq{hankellocalised} involving only moments up to $\mu_2^2$ are
\begin{gather}\label{hankellocalised22}
\left(\begin{array}{c|cc}
\mu_{0}^{0}& \mu_{1}^{0} & \mu_{1}^{1} \\\hline
\mu_{1}^{0}  &\mu_{2}^{0} & \mu_{2}^{1} \\
\mu_{1}^{1} & \mu_{2}^{1} & \mu_{2}^{2} 
\end{array}\right)\succ 0\,\quad
 \mu_{1}^{0} >0\,\quad\mu_{1}^{1} >0
\,\quad\mu_{2}^{1} >0\nn\\
 \mu_{0}^{0}-\bs^2 \mu_{1}^{0} >0\,\quad  \mu_{1}^{0}-\bs^2 \mu_{2}^{0} >0\,\quad  \mu_{1}^{1}-\bs^2 \mu_{2}^{1} >0\label{mu22bounds}
\end{gather}
and represent a simple (albeit not optimal) subset of all bounds.

\subsection{Bounds on Wilson Coefficients}
Now, near-forward bounds on Wilson coefficients stem from comparing, order by order in $t$,  IR and UV definitions of arcs in terms of moments. Using  \eq{arcsexplicit} and expanding the loop contribution \eq{deltaa0}, we find from $a_0(\bs,t)$,
\begin{gather}
g_{2,0}+\frac{\beta_4\bs^2}{2}=\mu_0^0\,,\quad -g_{3,1}-\frac{61}{42}\beta_4\bs=\mu_{1}^1-\frac{3}{2}\mu_{1}^0\,,\nn\\
2 g_{4,0}+\beta_4\left(2\log \bs+\frac{\log (m^2/\bs)}{42}\right)= \frac{\mu_{2}^2}{4} -2\mu_{2}^1 +2\mu_{2}^0\,.
\label{a0decomposed}
\end{gather}
The term proportional to $\log m^2$ represents the leading effect at finite mass, which acts here as a regulator for the otherwise  divergent expression as $t\rightarrow 0$.

From \eq{mu22bounds} (in particular $\mu_1^1>0$ and $\mu_0^0-\bs^2\mu_1^0>0$) we read  the upper bound 
\begin{equation}
g_{3,1}\bs<\frac{3}{2}g_{2,0}-\frac{10}{7}\beta_4\bs^2
\end{equation}
which for $\beta_4\to 0$ reduces to the tree-level values discussed in~\cite{deRham:2017avq,Bellazzini:2020cot}.

On the other hand the moments $\mu_1^1,\mu^2_2$ do not have upper bounds in   \eq{mu22bounds}, and consequently there appears to be no lower bound on $g_{3,1}$.\footnote{This is a consequence of the $\ell$ domain being non-compact:  $\mu_n^q$  can be larger and larger as $q$ increases. 
In contrast, $0\leq x\leq1$, and moments $\mu_n^q$ are monotonically decreasing in $n$.
} A lower bound stems from realizing that,  because of full $s-t-u$ crossing symmetry~\cite{Tolley:2020gtv,Caron-Huot:2021rmr},  $g_{4,0}$ appears also in  the second arc at $t=0$,
\begin{equation}
a_1=g_{4,0}+\beta_4\log \bs=\mu_2^0\,.
\end{equation}
Comparing this to the second line in \eq{a0decomposed} leads to a null constraint which, taking into account loop effects, reads,
\begin{equation}\label{nullconstfud}
\mu_2^2=8\mu_2^1+\frac{2\beta_4}{21}\log \frac{m^2}{\bs}\,.
\end{equation}
Null constraints relate higher and lower moments in $J^2$ and, together with bounds on moments, lead also to a lower bound on $g_{3,1}$~\cite{Caron-Huot:2020cmc,Tolley:2020gtv,Sinha:2020win,Chiang:2021ziz}. The simplest way to see this is to combine the null constraint with the condition $\mu_1^1-\bs^2\mu_2^1>0$ and $\mu_2^2\mu_0^0>(\mu_1^1)^2$ (the latter follows from positivity of the minors in the positive definite matrix of \eq{mu22bounds}), to obtain,
\begin{equation}\label{boundwithloopmoments}
g_{3,0}\bs>-4g_{2,0}\left(1+\sqrt{1-\frac{g_{2,0}\bs^2\log \frac{m^2}{\bs}}{240\times16\pi^2 }}\right)-\frac{10}{7}\beta_4\bs^2\,.
\end{equation}
In absence of loop effects, this reduces to $g_{3,0}\bs>-8g_{2,0}$; instead using all relations from \eq{mu22bounds} we find $g_{3,0}\bs>-6.5g_{2,0}$; finally, using moments up to $\mu_6^6$, we find $g_{3,0}\bs \gtrsim -5.3g_{2,0}$ in agreement with \cite{Caron-Huot:2020cmc,Tolley:2020gtv,Sinha:2020win,Chiang:2021ziz} -- this is an example of how the 2D moment problem converges as more and more moments are taken into account. 

\begin{figure}[t]\centering
\includegraphics[width=0.45\textwidth]{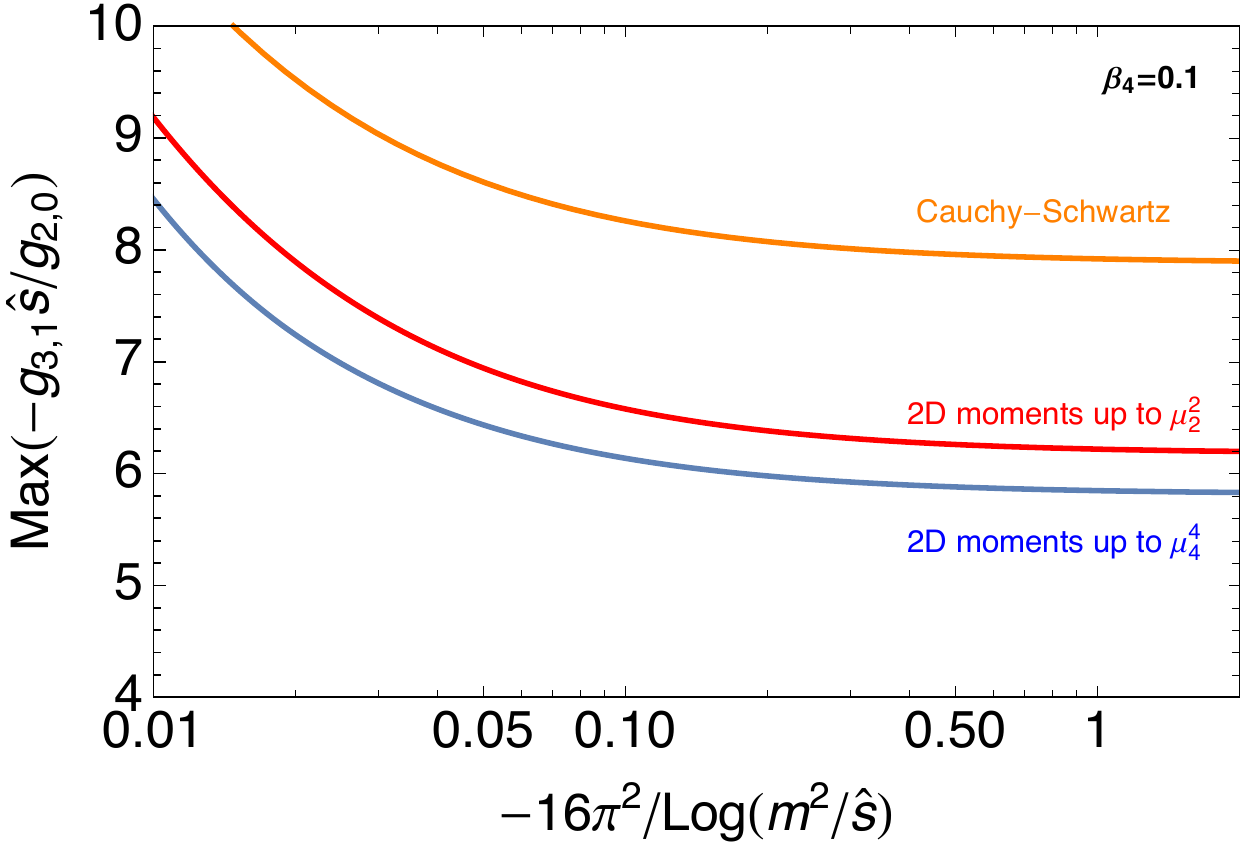}
\caption{{\it Bounds on ratio $g_{3,1}\bs/g_{2,0}$ as a function of the particle mass $m$, for fixed $\beta_4\bs^2=0.1$. Lower lying  curves correspond to bounds involving more moments in the UV: in orange only Cauchy-Schwartz \eq{boundwithloopmoments}, in red moments ut to $\mu_2^2$ as in \eq{mu22bounds}, and in blue moments up to $\mu_4^4$ (numerical). All curves diverge as $m\rightarrow 0$. }}\label{figm}
\end{figure}

We show these bounds, as function of $\log m^2\bs$, in Fig.~\ref{figm}.\footnote{At fixed $m$, with more moments, the lower bound improves, because the solution to the 2D moment problem is more precise, but it still diverges as $m\to 0$. In contrast to the tree-level limit, here  higher null constraints (from expressing any $g_{p,q}$  in terms of different combinations of moments, similarly to \eq{nullconstfud}) do not  improve the bound, because higher $t$-derivatives of  \eq{deltaa0} are more and more singular $\partial^k_t a_0\sim m^{-2(k-2)}$ as $m\to 0$.} The interesting feature is that in the exact $m=0$ limit, although the upper bound is untouched, the lower bound disappears completely.
 
 Nevertheless, for practical purposes, in the case of the Goldstone, as soon as the mass is finite the impact of loop effects is limited.
Indeed, even considering the most favorable phenomenological conditions for the EFT,  in which the cutoff is at the Planck scale $\bs=M_{pl}^2$, while the mass of the particle is at the lowest testable scale (Hubble) $m=H_0$, the logarithm is of order $\log \frac{m^2}{\bs}\approx - 280$  and the departures from the tree-level limit are of the size of a loop factor $\sim g_{2,0}^2s^2/16\pi^2$.

In the approach of this section,  positive functions are approximated by polynomials. This is a strength, because they provide a simple and systematic path to positivity.  At the same time, polynomials are the weak link of this approach, as they do not converge uniformly to continuum  functions on non-compact domains. For instance, consider the combination of moments associated with integrating $\cos J^2$ over the measure $d\mu$ in \eq{2dmoments}, $\mu^{cos}\equiv\sum_{n=0}^\infty(-1)^n\mu_0^{2n}/(2n!)$. Clearly $\mu^{cos}<\mu_0^{0}$, since $|\cos{J^2}|<1$; yet this feature is invisible at any finite order of moments.
In the next section we provide an alternative method to extract bounds, that attacks directly the integrand boundedness.

\section{Dispersion Relations at large~$-t$}

In this section, we assume   the amplitude to be analytic {in  the cut complex $s$ plane for fixed large negative $t$ (within the EFT validity)} -- as  often (though not always \cite{Guerrieri:2021tak}) assumed in the modern S-matrix bootstrap approach \cite{Paulos:2016fap,Paulos:2016but,Paulos:2017fhb}. This provides a way of regulating the singularities using finite $t$ rather than mass.
Analyticity in $s$ for finite $t<0$ has been rigorously proven~\cite{Bros:1965kbd}, except for a region of size $\sim(-t)^3$ around the origin. For $t$ within the range of EFT validity, the amplitude is explicitly calculable and known, and it displays  in fact no non-analyticity.

{The extended domain of analyticity enables us to study dispersion relations at fixed large negative $t$, without the need to expand them around $t\approx 0$. Unfortunately, }contrary to analogous quantities defined at~$t=0$, Legendre polynomials are not positive, and therefore the integrand \eq{archdeftUV} is \emph{not positive}. For $-\bs<t\leq0$, however, the integrand $I_t(\ell,\os)$ is \emph{bounded from above and below}, because the Legendre polynomials are themselves bounded for all even $\ell$ and $\os\in[\bs,\infty[$
\begin{equation}\label{Plbound}
-\frac{1}{2}\leq \min_{\ell,\os}P_\ell(1+\frac{2t}{\os\!\!-\!2m^2})\leq P_\ell(\cos\theta)\leq 1 \,.
\end{equation}
Since the integrand is bounded, we can pull it out of the integral and bound arcs $a_n(\bs,t')$ in terms of arcs at $t=0$ (where $P_\ell(\cos\theta)=P_\ell(1)=1$), 
\begin{equation}\label{a0}
a_n(\bs,0)= \frac{2}{\pi}\int_{\bs}^\infty \sum_{\ell=0}^\infty \frac{d\os\textrm{Im} \hat f_\ell(\os) }{\hat s^{\prime\,2n+3}}=\mu_{2n}^0\,,
\end{equation}
which, as shown above, are strictly positive  moments of a 1D distribution~\cite{Bellazzini:2020cot}.

Combining Eqs.~(\ref{archdeftUV}-\ref{a0}) we find the constraint,
\begin{equation}\label{anbracket}
\min_{\ell,\os}I^n_t(\ell,\os) \leq \frac{a_n(\bs,t)}{ a_n(\bs,0)}\leq \frac{\bs+\frac{t}{2} }{[\bs(\bs+t)]^{n+2}} \,.
\end{equation}
Here, for all $t$, the upper bound is exactly saturated in $\ell=0$ or $\os\to \infty$ (corresponding to $\cos\theta=1$), $P_0(\cos\theta)=P_\ell(1)=1$. Instead, for generic value of $t/\bs$ the lower bound is  determined by  different points in the $\ell,\os$ domain:
for $t/\bs\sim-1/2$, $\min_{\ell,\os}I_t(\ell,\os)$ is saturated by the $\ell=2$ polynomial, and as $t/\bs\to0$ or $1$ it is saturated by larger and larger values of $\ell$.

\begin{figure}[t]\centering
\includegraphics[width=0.45\textwidth]{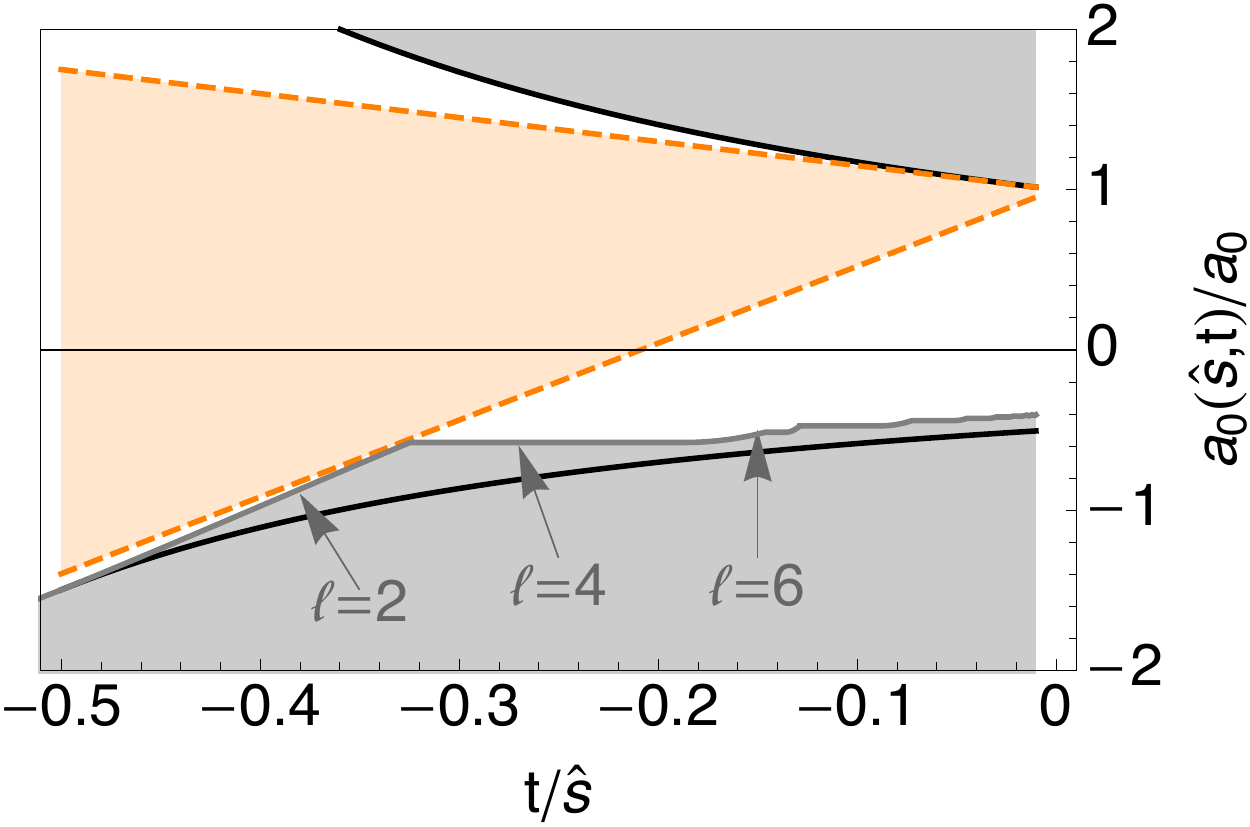}\hspace{1cm}
\caption{{\it Bounds on arc $a_0(\bs,t)$ as a function of $t$. In gray the region excluded by UV positivity bounds (in black the region excluded by the simple approximation on the extremes of \eq{Plbound}).
Orange dashed lines illustrate IR arcs calculated with the approximated EFT ($a_0(\bs,t)$ in \eq{arcsexplicit} truncated at $O(t)$), for extremal values of~$g_{3,1}$.}\label{figtrump}}
\end{figure}

The region excluded is illustrated for $n=0$ by the grey area in Fig.~\ref{figtrump} (the black lines correspond to the simple $t$-independent bound $-\frac{1}{2}\leq P_\ell(\cos\theta)\leq 1$).

\vspace{5mm}
We can now compare the arcs computed within the EFT \eq{arcsexplicit} with the arcs bounded by UV unitarity/causality in \eq{anbracket}. Calling $s_{max}$ the theory's cutoff, we will first consider the kinematics,
\begin{equation}\label{|smalltapp}
|t|\ll \bs \ll \bs_{max}\,.
\end{equation}
 The second inequality labels the regime where loop effects are under control, even in strongly coupled theories. 
 
 We first discuss the tree-level limit in which we neglect these effects altogether {such that the IR amplitude is well described by \eq{amptreeXY}}. We then invoke  the first inequality in \eq{|smalltapp} so that  the first arc is well approximated by, 
\begin{equation}\label{apparc}
a_0(\bs,t)\approx g_{2, 0} - t g_{3, 1}\,.
\end{equation}
and  higher-order terms can be neglected.\footnote{ While neglecting higher orders is a customary assumption in the context of EFTs, it is plausible that $g_{3,1}$ is not suppressed w.r.t. $g_{2,0}$ in units of $\bs$, while higher order terms are: a situation that corresponds  to a system with approximate Galileon symmetry). This is possibility that we are exploring under assumption~\eq{apparc}.}
 
 The approximate IR tree-level EFT arc is  illustrated   in Fig.~\ref{figtrump}, by lines with steepness $-\hat s g_{3, 1}/g_{2, 0}$: its extrema (orange dashed) are found by requiring that $a_0(\bs,t)$ lies within the UV allowed region \eq{anbracket}, for all values of~$-1\leq t/\bs\leq0$. 
Equating IR and  UV  arcs, and dividing by $a_0(\bs,0)$, we find,
\begin{equation}\label{boundformg3}
\frac{3}{2}>\bs \frac{g_{3,1}}{g_{2,0}}>-\min_{t}\left(\frac{\min_{\ell,\os}I^0_t-1}{-t}\right)\,.
\end{equation}
The upper bound is  saturated by the $\ell=0$, $\os=\bs$ UV contribution, corresponding to  a weakly coupled scalar with mass $M^2=\bs$. This bound comes from the near forward limit, where our assumption of neglecting higher order terms is exact. It can be found analytically by comparing $\partial|_{t=0}a_0(\bs,t)$  in the UV (the rhs of \eq{anbracket}) and the IR, \eq{apparc}, and is in fact equivalent to the moment problem approach.

The lower bound instead requires  $\min_{\ell,\os}I^{0}_t$, which implies finding the minima of (Legendre) polynomials, which is saturated by larger $\ell$ as $t/\bs\to0$. Nevertheless, the bound \eq{boundformg3} is dominated by finite $|t|/\bs$ where $\min_{\ell,\os}I^{0}_t$ is at  the interception of the $\ell=2$ and $\ell=4$ contributions, which can be found by solving a 4th order equation. It leads to, 
 \begin{equation}\label{boundssmallt}
\frac{\hat s g_{3, 1}}{g_{2, 0}}\gtrsim -4.9\,,
\end{equation}
and is dominated by the region $t\approx -0.3 \bs$, {which explains  why $s$-analyticity at large negative $-t\gg m^2$ is necessary.} Notice that this is a preliminary result that relies on neglecting higher order terms.
We will come back in the next section to a way (proposed in \cite{Caron-Huot:2021rmr}) to partially circumvent the question of higher order terms, at least at tree-level.

\begin{figure}[th]\centering
\includegraphics[width=0.45\textwidth]{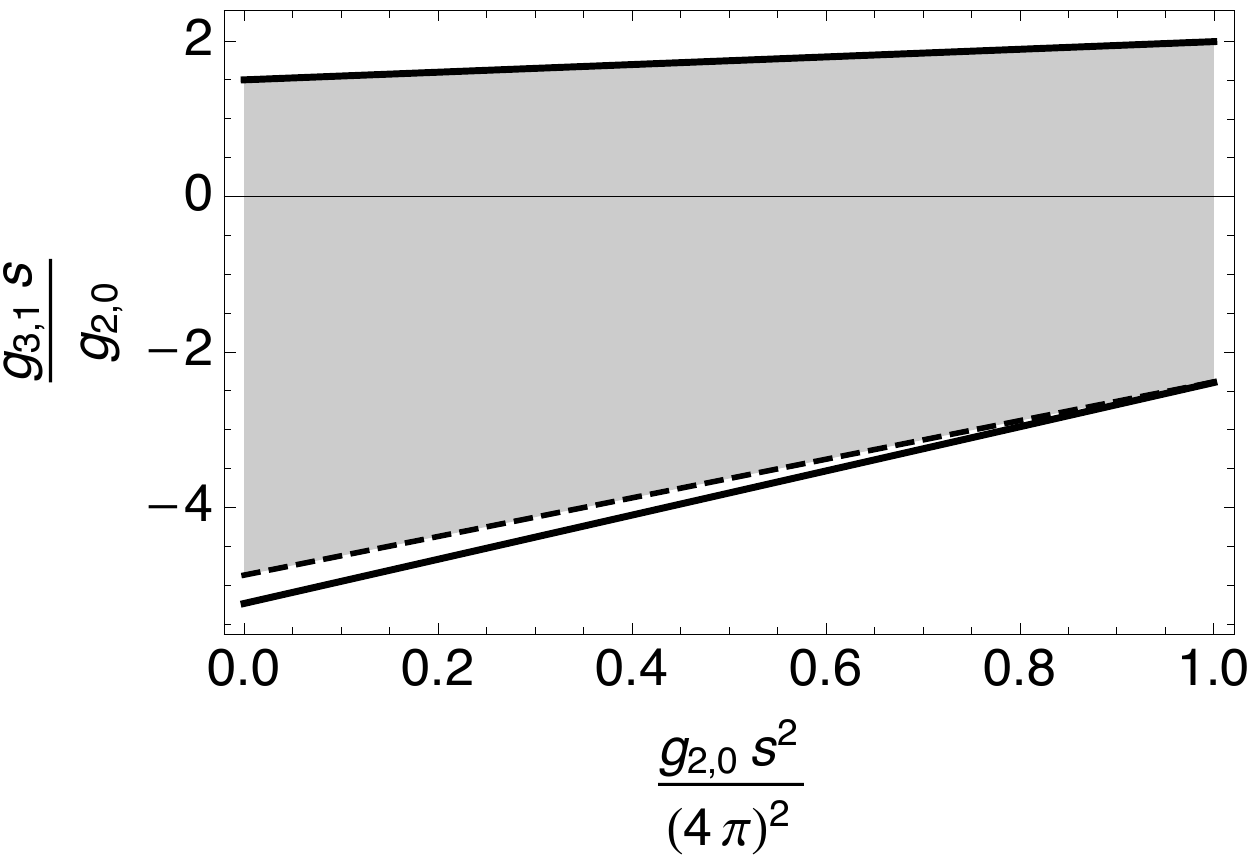}
\caption{{\it Bounds on the ratio $g_{3,1}\bs/g_{2,0}$ as a function of the coupling  $g_{2,0}\bs^2$. Grey area/dashed line: bounds neglecting terms $O(t^2)$ in $a_0(\bs,t)$ (but using the full loop contribution \eq{deltaa0}). Solid lines:  bounds from improved arcs, adding \eq{aimpIR} and \eq{deltaimpa0}.}}\label{figg}
\end{figure}

\vspace{5mm}
We now turn to loop effects, as captured by \eq{deltaa0}.
Despite its non-analyticity, both the amplitude and its first derivative remain finite at all $s,t,u$ within the EFT validity.
This is important, because the bounds presented in the previous paragraph depend  on the first derivative of $a_0(\bs,t)$ at $t=0$ and on the amplitude at finite $t\approx -0.3 \bs$, where loop effects have a finite impact. Ignoring terms at order $(t/\bs)^2\sim (0.3)^2$, loop effects are captured by the substitution, 
\begin{equation}
\frac{\hat s g_{3, 1}}{g_{2, 0}}\to \frac{\hat s g_{3,1} +\frac{61}{42}\beta_4\bs^2}{g_{2,0}+\frac{\beta_4\bs^2}{2}}
\end{equation}
in \eq{boundssmallt}. We illustrate this in Fig.~\ref{figg} (the area between the upper solid line and the lower dashed line). Contrary to bounds on moments, here loop effects have a finite impact on the lower bound, which survives also in the massless limit. In practice, the singularity is regulated, rather than by the mass, by the finite value of $t$ which happens to determine the lower bound.

\subsection{Resummed Higher Order Terms}
\begin{figure}[t]\centering
\includegraphics[width=0.45\textwidth]{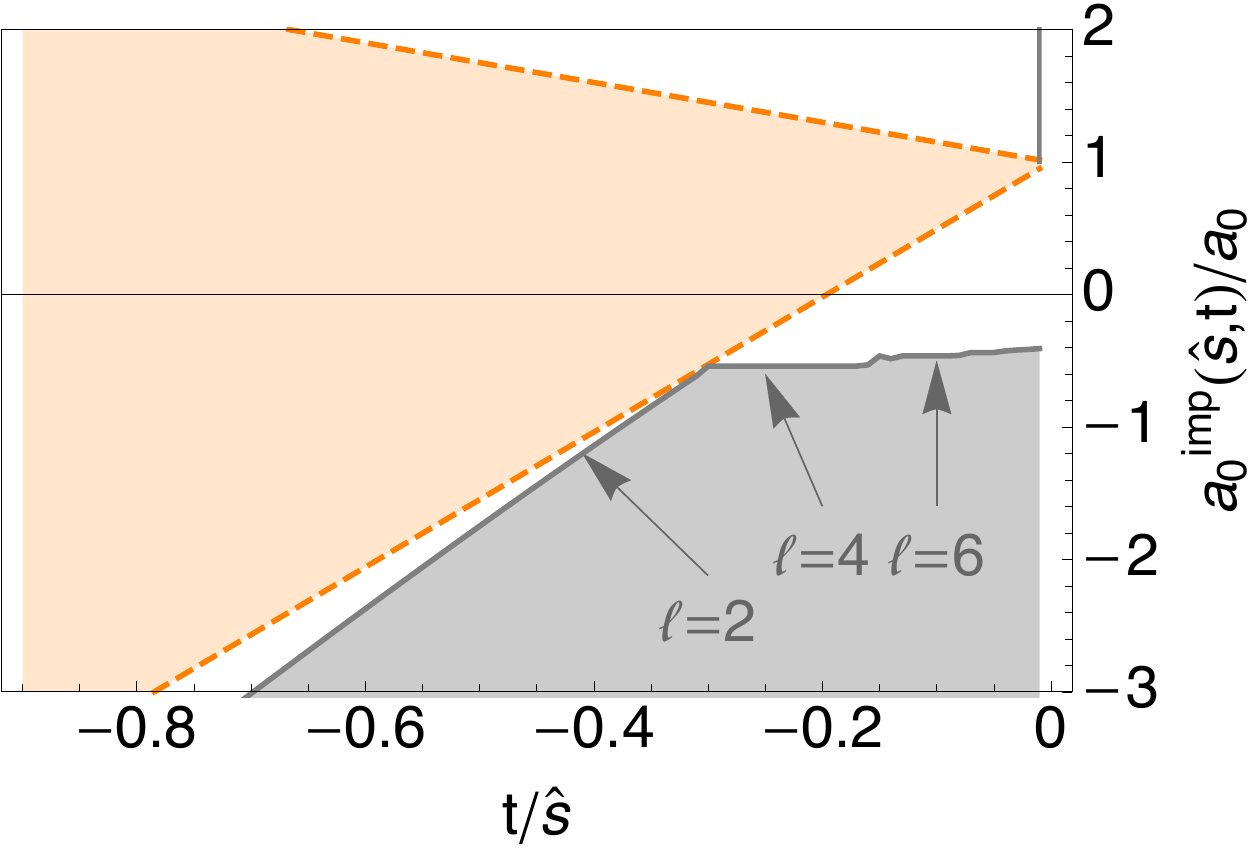}
\caption{{\it 
As in Fig.~\ref{figtrump}, but using  improved arcs $a^{imp}_0(\bs,t)$, and comparison with the now exact expression \eq{aimpIR} (orange) with extremal values given in  Eqs.~(\ref{boundexupper}).}\label{figtrumpImp}}
\end{figure} 

The previous paragraph relies on the approximation that at small $|t|\ll s$ the higher order terms  can  be neglected, see \eq{apparc}.
This assumption can be made rigorous using the near forward bounds from~\cite{Bellazzini:2020cot}, which imply that $0<g_{2n,0} \bs^{2n-2}<g_{2,0}$ and $(2n+1)g_{2,0}\bs^2/2>g_{2n+1,1}\bs^{2n+1}>-g_{3,1}\bs$.\footnote{\label{ftntBP}Notice that, even without crossing symmetry, terms that vanish in the forward amplitude still contribute at loop level to forward arcs, and are bounded in absolute value to be smaller than a loop factor times $g_{2,0}$~\cite{Nicolis:2009qm,Bellazzini:2017fep}.}
Alternatively, \emph{at tree-level}, higher Wilson coefficients can be eliminated altogether exploiting  crossing symmetry, via the ``improved'' arcs defined in Ref.~\cite{Caron-Huot:2021rmr}, leading to expressions that are valid for large angle scattering,
\begin{equation}\label{range}
-t\leq \bs \ll \bs_{max}\,,
\end{equation}
to be contrasted with \eq{|smalltapp}.

 Improved arcs rely on the full $s-t-u$ crossing symmetry of the amplitude to write terms of order $O(t^2)$ or higher as \emph{forward} higher arcs $a_{n\geq1}(\bs,0)$ or their first $t$-derivatives $\partial_t a_{n\geq1}(\bs,t)|_{t=0}$. These can then be  subtracted from both the IR and the UV representation of arcs. The resulting  improved arc~\cite{Caron-Huot:2021rmr},
\begin{equation}\label{improvedarc}
 a^{imp}_0=a_0(s,t)-\sum_{n\geq1}(n a_n-\partial_t a_n)|_{t=0}\,,
\end{equation}
 is defined in the IR and UV by,
\begin{equation}\label{aimpIR}
a^{imp}_0(\bs,t)=g_{2, 0} - t g_{3, 1}=\frac{2}{\pi}\int_{\bs}^\infty \!\!\sum_{\ell=0}^\infty \frac{d\os\textrm{Im} \hat f_\ell(\os)}{\hat s^{\prime3}}I^{imp}_t\,,
\end{equation}
with
\begin{equation}
I^{imp}_t\equiv \frac{(2\os+t)P_\ell(\cos\theta)}{2(\os+t)^{2}}-t^2\left(\frac{3 t+4\os}{2 (\os+t)^2}+\frac{\ell(\ell+1)t}{\hat s^{\prime 2}- t^2}\right)
\end{equation}
where $\ell(\ell+1)$ stems from the derivative of $P_\ell$ in the forward limit. 

We can then proceed as before: for fixed values of $-\bs \leq t\leq0$ we can find the extrema of $I^{imp}_t$ in $\os$ and~$\ell$. 
The maximum of $I^{imp}_t$ is discontinuous in $t$, since for $t<0$ it is dominated by the (positive) term proportional to $\ell(\ell+1)$, which diverges as $\ell\to \infty$; at $t=0$ instead, it is finite and has slope $-3/2$ (saturated by $\ell=0$ and $\os=\bs$). 
The minimum of $I^{imp}_t$, similarly to the previous paragraph, is dominated in the interesting region by the interplay of the $\ell=2$ and $\ell=4$ polynomials. The result is illustrated in  Fig.~\ref{figtrumpImp}: improved UV arcs must lie within the allowed (non-gray) region, for all values of $t$.

At tree-level, a comparison  with the (now) exact IR arc \eq{aimpIR} leads to
\begin{equation}\label{boundexupper}
\frac{3}{2}>\hat s \,\frac{g_{3, 1}}{g_{2, 0}}\gtrsim -5.18 \,\,,
\end{equation}
compatibly with \cite{Caron-Huot:2020cmc,Tolley:2020gtv,Caron-Huot:2021rmr,Chiang:2021ziz}.\footnote{In this approach we have not included higher null constraints.}
Notice that this bound appears weaker than the non-improved  one \eq{boundssmallt}, which ignored $O(t^2)$ terms.

The tree-level improved bound appears to be somewhat sharper, since higher-order tree-level Wilson coefficients have been eliminated exactly from both IR and UV parts of the dispersion relation. Once IR effects are taken into account, however, these higher-order Wilson coefficients reappear into arcs.
For instance, while the term $\propto g_{4,0}$ is canceled from $a_0(\bs,t)-2t^2a_1(\bs,0)-3t^4a_2(\bs,0)-\cdots$, there is a loop effect $\propto g_{2,0}g_{4,0}/16\pi^2$  which does not cancel. It is not possible to eliminate $g_{4,0}$ altogether from $a_0(\bs,t)$. 

So, contrary to tree-level arcs that could be improved into compact expressions involving a finite number of Wilson coefficients, loop level effects propagate all order coefficients  into arcs.
We will have to assume that these terms are small for all values of $s$ at which we evaluate bounds. We express this assumption as $\bs\ll \bs_{max}$ in \eq{range}.

Under this assumption, we can focus on the most relevant term, discussed already in \eq{amploop}, and compute the improved arc using the algorithm \eq{improvedarc}. In addition to the first expression in \eq{aimpIR}, we find,
\begin{equation}\label{deltaimpa0}
\delta a_0^{imp}=\frac{g_{2,0}^2}{16\pi^2}\left( \frac{-21 s^2+61 ts+t^2\log \left(1-\frac{s}{t}\right)}{60}-\frac{2t^3}{3s}\right) \,.
\end{equation}
instead of the non-improved version~\eq{deltaa0}. Including this term in the EFT, we can rederive the bound \eq{boundexupper}, which we show as the solid line in Fig.~\ref{figtrumpImp}. The improved result is very similar to the approximated one.

Importantly, the improvement algorithm \eq{improvedarc} involves higher arcs and their first derivatives at $t\to 0$. As argued in Eq.~(\ref{deltaa0}), these are finite, despite being evaluated in the forward kinematics. This result is  not a priori obvious. Moreover, thank to \eq{singha}, it  extends also to improvements of higher arcs. Indeed, consider   for example  $g_{6,2}$. The necessary improvement (the analog of \eq{aimpIR}) to eliminate all terms $g_{n,2}$, $n>6$ from $a_1(\bs,t)$, involves now second derivatives of higher arcs at $t=0$, $\partial^2_ta_n(\bs,t)|_{t=0,n\geq2}$. \eq{singha} implies that these are all regular. More generally, terms $\propto (stu)^q$ in \eq{amptreeXY}, appear first in arc $a_{q-1}(\bs,t)$ (see footnote \ref{footnote1}). Its improved version requires $\partial^q|_{t=0} a_n(\bs,t),\, n\geq q$, which is regular according to \eq{singha}. 

These results give strength to the methods pioneered in~\cite{Caron-Huot:2021rmr}: although designed for tree-level amplitudes, they hold also at loop-level for the simple scalar theory with no gravity we have considered in this work.

\section{Conclusions and Outlook}

We have studied the impact of calculable IR loop effects on EFT positivity bounds, focusing on the effect of IR singularities in the massless limit. This largely extends the initial investigations in \cite{Bellazzini:2020cot,Arkani-Hamed:2020blm} which focused  on  IR-finite contributions as  running effects.  
The role of IR-sensitive loop correction studied in the present work is actually more relevant for bounds relying on full crossing symmetry of the amplitude \cite{Tolley:2020gtv,Caron-Huot:2020cmc}, such as the lower bound on the coefficient ratio $g_{3,1}\bs/g_{2,0}$.
The singular behavior stems from a tension between  the need of probing the EFT at large enough energies~$\bs$, where bounds on Wilson coefficients are stronger, and small enough $t$ where the amplitude is analytic.

If $s$-analyticity of the amplitude is  granted only for a limited range of values taken by $t$,  $|t|\lesssim m^2$, we have shown how to extract the bounds, extending the approach of \cite{Bellazzini:2020cot} to the 2D moment problem, as in \cite{Chiang:2021ziz}. One-loop effects involving marginal couplings $\lambda \phi^4$ have no qualitative impact on the bounds.  On the other hand, loops involving the coupling $g_{2,0}$ induce a  $\log m^2/\bs$ divergence in the derivative of the arcs, which invalidates the bound for $m\to0$. Since  $g_{2,0}$ and $g_{3,1}$ do not run, the divergence is physical.

\vspace{1mm}

If instead the amplitude is analytic also for large negative $t$, then all the bounds are robust. We have shown this by introducing a novel and simple analytic understanding of dispersion relations at finite $t$ (which complement the numerical approach of \cite{Caron-Huot:2021rmr}), represented by the \emph{trumpet} in Fig.~\ref{figtrump}, which can be applied to any amplitude regular in the forward limit. In this approach, the upper bound on $g_{3,1}$ is shown to be dominated by the dispersion relation at $t/\bs\approx -0.3$, where the above singularities are absent. Moreover, this approach makes little use of crossing symmetry, {whose practical purpose is to guarantee that the higher order coefficients $g_{n,1} (n>4)$ are bounded by the lower coefficients.  
 It would be interesting to bring this analytic approach into a more systematic tool to approach all bounds. 

Moreover, the improvement procedure that removes higher order terms, still relies on evaluating infinitely many dispersion relations in the forward limit. 
We have studied the structure of all 1-loop effects in the scalar theory and found that, interestingly, all the necessary forward limits are regular.
It would be interesting to repeat this analysis in more complex theories, including $\phi^3$, flavor or gravity, to establish the robustness of this mixed forward/non-forward procedure.
Moreover, the study of higher loops of more relevant couplings as well theories with exactly massless particles, such as Yang-Mills theory, might reveal more singular behaviors.
Overall, it would be useful to develop an improvement procedure without the forward limit.

\section*{Acknowledgments}
We are very grateful to J. Elias-Miro for collaboration in the earl stages of this work, and we thank R.~Rattazzi and A.~Zhiboedov for discussions.
F.R. is particularly grateful  to J.B.~Lasserre, for numerous exchanges on the higher dimensional moment problem. 
The work of FR  is supported by the Swiss National Science
Foundation under grant no. PP00P2-170578.
\appendix

\section{Discrete Moments}\label{app:bf}

Treating the distribution in $J^2$ as continuous gives a convenient and manageable way to setting conservative constraints on the arcs without having to truncate the expansion in Legendre polynomials $P_\ell$ up to some $\ell_{max}$. However, we remark that it excludes polynomials that might be negative between integer values of $\ell$, so the bounds are not optimal. In \cite{Chiang:2021ziz} the condition is imposed via a set of positive determinants. In this Appendix propose a simple way to systematically improve the bounds.

The two-dimensional moments
\be
\mu_n^j = \frac{2}{\pi}\int_s^\infty \frac{ds}{s}\text{Im}f_\ell(s)\sum_\ell\frac{(\ell(\ell+1))^j}{s^{n+2}}
\ee
can be split as a sum of two terms
\bea
\label{eq:app2dmoments}
\mu_n^j &=&\sum_\ell^{L-1} (\ell(\ell+1))^j\frac{2}{\pi}\int_s^\infty \frac{ds}{s}\text{Im}f_\ell(s)\frac{1}{s^{n+2}}\\
&+&
\frac{2}{\pi}\int_s^\infty \frac{ds}{s}\int_{L(L+1)}^\infty f(J^2,s)\frac{(\ell(\ell+1))^j}{s^{n+2}}
\eea
In the first term, we are considering the two dimensional moment problem in a grid of $L$ sites, each site being a one-dimensional moment problems with a different measure $\text{Im}f_\ell(s)$. This automatically imposes the constraints for all moments above $j>L$. 

The grid in $J^2$ is specified by the zeroes of 
\be
\label{eq:appgrid}
g(J^2) = \prod_{i=0}^{L-1}(J^2-i(i+1))
\ee
and gives a relation between $L$ different moments. Explicitly, for $L=2$, $g(J^2)=J^2(J^2-2)$ and $\mu_n^{j+2}-2\mu_n^{j+1}=0$, so there are only two ($L$, in general) independent moments in $j$, $\mu_n^0$ and $\mu_n^1$. Writing the two-dimensional problem as continuous and then imposing the grid constraints in \eq{eq:appgrid} is equivalent of considering $L$ different one-dimensional problems. We find more convenient using the latter since, besides being more physical, one can impose the upper bound on the partial waves $\text{Im}f_\ell\leq 16\pi(2\ell+1)$,
\be
\frac{2}{\pi}\int_s^\infty \frac{ds}{s}\frac{\text{Im}f_\ell(s)}{s^{n+2}} \,<\, \frac{1}{s^{n+2}}\frac{32(2\ell+1)}{n+2}.
\ee

The second term in \eq{eq:app2dmoments} is the two-dimensional moment problem already described in the main text. Notice that now the domain is given by
\begin{equation}
x\geq0\,,\quad 1-x\geq0\,,\quad J^2-L(L+1)\geq 0\,.
\end{equation}
instead of \eq{boundariesSchmudigen}. This maintains information for asymptotic partial waves, and the approximation of not considering negative polynomials between $\ell$ above $L$ is extremely good already for relatively low $L\sim 4$ in the examples considered if numerical precision is desired, but setting $L=0$ as done in the main text allows for a simple and accurate analytical understanding.

\bibliography{bibs} 

\begin{thebibliography}{61}%
\makeatletter
\providecommand \@ifxundefined [1]{%
 \@ifx{#1\undefined}
}%
\providecommand \@ifnum [1]{%
 \ifnum #1\expandafter \@firstoftwo
 \else \expandafter \@secondoftwo
 \fi
}%
\providecommand \@ifx [1]{%
 \ifx #1\expandafter \@firstoftwo
 \else \expandafter \@secondoftwo
 \fi
}%
\providecommand \natexlab [1]{#1}%
\providecommand \enquote  [1]{``#1''}%
\providecommand \bibnamefont  [1]{#1}%
\providecommand \bibfnamefont [1]{#1}%
\providecommand \citenamefont [1]{#1}%
\providecommand \href@noop [0]{\@secondoftwo}%
\providecommand \href [0]{\begingroup \@sanitize@url \@href}%
\providecommand \@href[1]{\@@startlink{#1}\@@href}%
\providecommand \@@href[1]{\endgroup#1\@@endlink}%
\providecommand \@sanitize@url [0]{\catcode `\\12\catcode `\$12\catcode
  `\&12\catcode `\#12\catcode `\^12\catcode `\_12\catcode `\%12\relax}%
\providecommand \@@startlink[1]{}%
\providecommand \@@endlink[0]{}%
\providecommand \url  [0]{\begingroup\@sanitize@url \@url }%
\providecommand \@url [1]{\endgroup\@href {#1}{\urlprefix }}%
\providecommand \urlprefix  [0]{URL }%
\providecommand \Eprint [0]{\href }%
\providecommand \doibase [0]{http://dx.doi.org/}%
\providecommand \selectlanguage [0]{\@gobble}%
\providecommand \bibinfo  [0]{\@secondoftwo}%
\providecommand \bibfield  [0]{\@secondoftwo}%
\providecommand \translation [1]{[#1]}%
\providecommand \BibitemOpen [0]{}%
\providecommand \bibitemStop [0]{}%
\providecommand \bibitemNoStop [0]{.\EOS\space}%
\providecommand \EOS [0]{\spacefactor3000\relax}%
\providecommand \BibitemShut  [1]{\csname bibitem#1\endcsname}%
\let\auto@bib@innerbib\@empty
\bibitem [{\citenamefont {Pham}\ and\ \citenamefont
  {Truong}(1985)}]{Pham:1985cr}%
  \BibitemOpen
  \bibfield  {author} {\bibinfo {author} {\bibfnamefont {T.~N.}\ \bibnamefont
  {Pham}}\ and\ \bibinfo {author} {\bibfnamefont {T.~N.}\ \bibnamefont
  {Truong}},\ }\href {\doibase 10.1103/PhysRevD.31.3027} {\bibfield  {journal}
  {\bibinfo  {journal} {Phys. Rev.}\ }\textbf {\bibinfo {volume} {D31}},\
  \bibinfo {pages} {3027} (\bibinfo {year} {1985})}\BibitemShut {NoStop}%
\bibitem [{\citenamefont {Ananthanarayan}\ \emph {et~al.}(1995)\citenamefont
  {Ananthanarayan}, \citenamefont {Toublan},\ and\ \citenamefont
  {Wanders}}]{Ananthanarayan:1994hf}%
  \BibitemOpen
  \bibfield  {author} {\bibinfo {author} {\bibfnamefont {B.}~\bibnamefont
  {Ananthanarayan}}, \bibinfo {author} {\bibfnamefont {D.}~\bibnamefont
  {Toublan}}, \ and\ \bibinfo {author} {\bibfnamefont {G.}~\bibnamefont
  {Wanders}},\ }\href {\doibase 10.1103/PhysRevD.51.1093} {\bibfield  {journal}
  {\bibinfo  {journal} {Phys. Rev. D}\ }\textbf {\bibinfo {volume} {51}},\
  \bibinfo {pages} {1093} (\bibinfo {year} {1995})},\ \Eprint
  {http://arxiv.org/abs/hep-ph/9410302} {arXiv:hep-ph/9410302} \BibitemShut
  {NoStop}%
\bibitem [{\citenamefont {Pennington}\ and\ \citenamefont
  {Portoles}(1995)}]{Pennington:1994kc}%
  \BibitemOpen
  \bibfield  {author} {\bibinfo {author} {\bibfnamefont {M.~R.}\ \bibnamefont
  {Pennington}}\ and\ \bibinfo {author} {\bibfnamefont {J.}~\bibnamefont
  {Portoles}},\ }\href {\doibase 10.1016/0370-2693(94)01551-M} {\bibfield
  {journal} {\bibinfo  {journal} {Phys. Lett. B}\ }\textbf {\bibinfo {volume}
  {344}},\ \bibinfo {pages} {399} (\bibinfo {year} {1995})},\ \Eprint
  {http://arxiv.org/abs/hep-ph/9409426} {arXiv:hep-ph/9409426} \BibitemShut
  {NoStop}%
\bibitem [{\citenamefont {Adams}\ \emph {et~al.}(2006)\citenamefont {Adams},
  \citenamefont {Arkani-Hamed}, \citenamefont {Dubovsky}, \citenamefont
  {Nicolis},\ and\ \citenamefont {Rattazzi}}]{Adams:2006sv}%
  \BibitemOpen
  \bibfield  {author} {\bibinfo {author} {\bibfnamefont {A.}~\bibnamefont
  {Adams}}, \bibinfo {author} {\bibfnamefont {N.}~\bibnamefont {Arkani-Hamed}},
  \bibinfo {author} {\bibfnamefont {S.}~\bibnamefont {Dubovsky}}, \bibinfo
  {author} {\bibfnamefont {A.}~\bibnamefont {Nicolis}}, \ and\ \bibinfo
  {author} {\bibfnamefont {R.}~\bibnamefont {Rattazzi}},\ }\href {\doibase
  10.1088/1126-6708/2006/10/014} {\bibfield  {journal} {\bibinfo  {journal}
  {JHEP}\ }\textbf {\bibinfo {volume} {10}},\ \bibinfo {pages} {014} (\bibinfo
  {year} {2006})},\ \Eprint {http://arxiv.org/abs/hep-th/0602178}
  {arXiv:hep-th/0602178 [hep-th]} \BibitemShut {NoStop}%
\bibitem [{\citenamefont {Bellazzini}(2017)}]{Bellazzini:2016xrt}%
  \BibitemOpen
  \bibfield  {author} {\bibinfo {author} {\bibfnamefont {B.}~\bibnamefont
  {Bellazzini}},\ }\href {\doibase 10.1007/JHEP02(2017)034} {\bibfield
  {journal} {\bibinfo  {journal} {JHEP}\ }\textbf {\bibinfo {volume} {02}},\
  \bibinfo {pages} {034} (\bibinfo {year} {2017})},\ \Eprint
  {http://arxiv.org/abs/1605.06111} {arXiv:1605.06111 [hep-th]} \BibitemShut
  {NoStop}%
\bibitem [{\citenamefont {de~Rham}\ \emph
  {et~al.}(2018{\natexlab{a}})\citenamefont {de~Rham}, \citenamefont
  {Melville}, \citenamefont {Tolley},\ and\ \citenamefont
  {Zhou}}]{deRham:2017zjm}%
  \BibitemOpen
  \bibfield  {author} {\bibinfo {author} {\bibfnamefont {C.}~\bibnamefont
  {de~Rham}}, \bibinfo {author} {\bibfnamefont {S.}~\bibnamefont {Melville}},
  \bibinfo {author} {\bibfnamefont {A.~J.}\ \bibnamefont {Tolley}}, \ and\
  \bibinfo {author} {\bibfnamefont {S.-Y.}\ \bibnamefont {Zhou}},\ }\href
  {\doibase 10.1007/JHEP03(2018)011} {\bibfield  {journal} {\bibinfo  {journal}
  {JHEP}\ }\textbf {\bibinfo {volume} {03}},\ \bibinfo {pages} {011} (\bibinfo
  {year} {2018}{\natexlab{a}})},\ \Eprint {http://arxiv.org/abs/1706.02712}
  {arXiv:1706.02712 [hep-th]} \BibitemShut {NoStop}%
\bibitem [{\citenamefont {Vecchi}(2007)}]{Vecchi:2007na}%
  \BibitemOpen
  \bibfield  {author} {\bibinfo {author} {\bibfnamefont {L.}~\bibnamefont
  {Vecchi}},\ }\href {\doibase 10.1088/1126-6708/2007/11/054} {\bibfield
  {journal} {\bibinfo  {journal} {JHEP}\ }\textbf {\bibinfo {volume} {11}},\
  \bibinfo {pages} {054} (\bibinfo {year} {2007})},\ \Eprint
  {http://arxiv.org/abs/0704.1900} {arXiv:0704.1900 [hep-ph]} \BibitemShut
  {NoStop}%
\bibitem [{\citenamefont {Nicolis}\ \emph {et~al.}(2010)\citenamefont
  {Nicolis}, \citenamefont {Rattazzi},\ and\ \citenamefont
  {Trincherini}}]{Nicolis:2009qm}%
  \BibitemOpen
  \bibfield  {author} {\bibinfo {author} {\bibfnamefont {A.}~\bibnamefont
  {Nicolis}}, \bibinfo {author} {\bibfnamefont {R.}~\bibnamefont {Rattazzi}}, \
  and\ \bibinfo {author} {\bibfnamefont {E.}~\bibnamefont {Trincherini}},\
  }\href {\doibase 10.1007/JHEP05(2010)095, 10.1007/JHEP11(2011)128} {\bibfield
   {journal} {\bibinfo  {journal} {JHEP}\ }\textbf {\bibinfo {volume} {05}},\
  \bibinfo {pages} {095} (\bibinfo {year} {2010})},\ \bibinfo {note} {[Erratum:
  JHEP11,128(2011)]},\ \Eprint {http://arxiv.org/abs/0912.4258}
  {arXiv:0912.4258 [hep-th]} \BibitemShut {NoStop}%
\bibitem [{\citenamefont {Bellazzini}\ \emph {et~al.}(2014)\citenamefont
  {Bellazzini}, \citenamefont {Martucci},\ and\ \citenamefont
  {Torre}}]{Bellazzini:2014waa}%
  \BibitemOpen
  \bibfield  {author} {\bibinfo {author} {\bibfnamefont {B.}~\bibnamefont
  {Bellazzini}}, \bibinfo {author} {\bibfnamefont {L.}~\bibnamefont
  {Martucci}}, \ and\ \bibinfo {author} {\bibfnamefont {R.}~\bibnamefont
  {Torre}},\ }\href {\doibase 10.1007/JHEP09(2014)100} {\bibfield  {journal}
  {\bibinfo  {journal} {JHEP}\ }\textbf {\bibinfo {volume} {09}},\ \bibinfo
  {pages} {100} (\bibinfo {year} {2014})},\ \Eprint
  {http://arxiv.org/abs/1405.2960} {arXiv:1405.2960 [hep-th]} \BibitemShut
  {NoStop}%
\bibitem [{\citenamefont {de~Rham}\ \emph {et~al.}(2017)\citenamefont
  {de~Rham}, \citenamefont {Melville}, \citenamefont {Tolley},\ and\
  \citenamefont {Zhou}}]{deRham:2017avq}%
  \BibitemOpen
  \bibfield  {author} {\bibinfo {author} {\bibfnamefont {C.}~\bibnamefont
  {de~Rham}}, \bibinfo {author} {\bibfnamefont {S.}~\bibnamefont {Melville}},
  \bibinfo {author} {\bibfnamefont {A.~J.}\ \bibnamefont {Tolley}}, \ and\
  \bibinfo {author} {\bibfnamefont {S.-Y.}\ \bibnamefont {Zhou}},\ }\href
  {\doibase 10.1103/PhysRevD.96.081702} {\bibfield  {journal} {\bibinfo
  {journal} {Phys. Rev.}\ }\textbf {\bibinfo {volume} {D96}},\ \bibinfo {pages}
  {081702} (\bibinfo {year} {2017})},\ \Eprint
  {http://arxiv.org/abs/1702.06134} {arXiv:1702.06134 [hep-th]} \BibitemShut
  {NoStop}%
\bibitem [{\citenamefont {Distler}\ \emph {et~al.}(2007)\citenamefont
  {Distler}, \citenamefont {Grinstein}, \citenamefont {Porto},\ and\
  \citenamefont {Rothstein}}]{Distler:2006if}%
  \BibitemOpen
  \bibfield  {author} {\bibinfo {author} {\bibfnamefont {J.}~\bibnamefont
  {Distler}}, \bibinfo {author} {\bibfnamefont {B.}~\bibnamefont {Grinstein}},
  \bibinfo {author} {\bibfnamefont {R.~A.}\ \bibnamefont {Porto}}, \ and\
  \bibinfo {author} {\bibfnamefont {I.~Z.}\ \bibnamefont {Rothstein}},\ }\href
  {\doibase 10.1103/PhysRevLett.98.041601} {\bibfield  {journal} {\bibinfo
  {journal} {Phys. Rev. Lett.}\ }\textbf {\bibinfo {volume} {98}},\ \bibinfo
  {pages} {041601} (\bibinfo {year} {2007})},\ \Eprint
  {http://arxiv.org/abs/hep-ph/0604255} {arXiv:hep-ph/0604255 [hep-ph]}
  \BibitemShut {NoStop}%
\bibitem [{\citenamefont {Manohar}\ and\ \citenamefont
  {Mateu}(2008)}]{Manohar:2008tc}%
  \BibitemOpen
  \bibfield  {author} {\bibinfo {author} {\bibfnamefont {A.~V.}\ \bibnamefont
  {Manohar}}\ and\ \bibinfo {author} {\bibfnamefont {V.}~\bibnamefont
  {Mateu}},\ }\href {\doibase 10.1103/PhysRevD.77.094019} {\bibfield  {journal}
  {\bibinfo  {journal} {Phys. Rev.}\ }\textbf {\bibinfo {volume} {D77}},\
  \bibinfo {pages} {094019} (\bibinfo {year} {2008})},\ \Eprint
  {http://arxiv.org/abs/0801.3222} {arXiv:0801.3222 [hep-ph]} \BibitemShut
  {NoStop}%
\bibitem [{\citenamefont {Low}\ \emph {et~al.}(2010)\citenamefont {Low},
  \citenamefont {Rattazzi},\ and\ \citenamefont {Vichi}}]{Low:2009di}%
  \BibitemOpen
  \bibfield  {author} {\bibinfo {author} {\bibfnamefont {I.}~\bibnamefont
  {Low}}, \bibinfo {author} {\bibfnamefont {R.}~\bibnamefont {Rattazzi}}, \
  and\ \bibinfo {author} {\bibfnamefont {A.}~\bibnamefont {Vichi}},\ }\href
  {\doibase 10.1007/JHEP04(2010)126} {\bibfield  {journal} {\bibinfo  {journal}
  {JHEP}\ }\textbf {\bibinfo {volume} {1004}},\ \bibinfo {pages} {126}
  (\bibinfo {year} {2010})},\ \Eprint {http://arxiv.org/abs/0907.5413}
  {arXiv:0907.5413 [hep-ph]} \BibitemShut {NoStop}%
\bibitem [{\citenamefont {Falkowski}\ \emph {et~al.}(2012)\citenamefont
  {Falkowski}, \citenamefont {Rychkov},\ and\ \citenamefont
  {Urbano}}]{Falkowski:2012vh}%
  \BibitemOpen
  \bibfield  {author} {\bibinfo {author} {\bibfnamefont {A.}~\bibnamefont
  {Falkowski}}, \bibinfo {author} {\bibfnamefont {S.}~\bibnamefont {Rychkov}},
  \ and\ \bibinfo {author} {\bibfnamefont {A.}~\bibnamefont {Urbano}},\ }\href
  {\doibase 10.1007/JHEP04(2012)073} {\bibfield  {journal} {\bibinfo  {journal}
  {JHEP}\ }\textbf {\bibinfo {volume} {04}},\ \bibinfo {pages} {073} (\bibinfo
  {year} {2012})},\ \Eprint {http://arxiv.org/abs/1202.1532} {arXiv:1202.1532
  [hep-ph]} \BibitemShut {NoStop}%
\bibitem [{\citenamefont {Bellazzini}\ \emph {et~al.}(2017)\citenamefont
  {Bellazzini}, \citenamefont {Riva}, \citenamefont {Serra},\ and\
  \citenamefont {Sgarlata}}]{Bellazzini:2017bkb}%
  \BibitemOpen
  \bibfield  {author} {\bibinfo {author} {\bibfnamefont {B.}~\bibnamefont
  {Bellazzini}}, \bibinfo {author} {\bibfnamefont {F.}~\bibnamefont {Riva}},
  \bibinfo {author} {\bibfnamefont {J.}~\bibnamefont {Serra}}, \ and\ \bibinfo
  {author} {\bibfnamefont {F.}~\bibnamefont {Sgarlata}},\ }\href {\doibase
  10.1007/JHEP11(2017)020} {\bibfield  {journal} {\bibinfo  {journal} {JHEP}\
  }\textbf {\bibinfo {volume} {11}},\ \bibinfo {pages} {020} (\bibinfo {year}
  {2017})},\ \Eprint {http://arxiv.org/abs/1706.03070} {arXiv:1706.03070
  [hep-ph]} \BibitemShut {NoStop}%
\bibitem [{\citenamefont {Bellazzini}\ and\ \citenamefont
  {Riva}(2018)}]{Bellazzini:2018paj}%
  \BibitemOpen
  \bibfield  {author} {\bibinfo {author} {\bibfnamefont {B.}~\bibnamefont
  {Bellazzini}}\ and\ \bibinfo {author} {\bibfnamefont {F.}~\bibnamefont
  {Riva}},\ }\href {\doibase 10.1103/PhysRevD.98.095021} {\bibfield  {journal}
  {\bibinfo  {journal} {Phys. Rev.}\ }\textbf {\bibinfo {volume} {D98}},\
  \bibinfo {pages} {095021} (\bibinfo {year} {2018})},\ \Eprint
  {http://arxiv.org/abs/1806.09640} {arXiv:1806.09640 [hep-ph]} \BibitemShut
  {NoStop}%
\bibitem [{\citenamefont {Zhang}\ and\ \citenamefont
  {Zhou}(2019)}]{Zhang:2018shp}%
  \BibitemOpen
  \bibfield  {author} {\bibinfo {author} {\bibfnamefont {C.}~\bibnamefont
  {Zhang}}\ and\ \bibinfo {author} {\bibfnamefont {S.-Y.}\ \bibnamefont
  {Zhou}},\ }\href {\doibase 10.1103/PhysRevD.100.095003} {\bibfield  {journal}
  {\bibinfo  {journal} {Phys. Rev.}\ }\textbf {\bibinfo {volume} {D100}},\
  \bibinfo {pages} {095003} (\bibinfo {year} {2019})},\ \Eprint
  {http://arxiv.org/abs/1808.00010} {arXiv:1808.00010 [hep-ph]} \BibitemShut
  {NoStop}%
\bibitem [{\citenamefont {Remmen}\ and\ \citenamefont
  {Rodd}(2019)}]{Remmen:2019cyz}%
  \BibitemOpen
  \bibfield  {author} {\bibinfo {author} {\bibfnamefont {G.~N.}\ \bibnamefont
  {Remmen}}\ and\ \bibinfo {author} {\bibfnamefont {N.~L.}\ \bibnamefont
  {Rodd}},\ }\href {\doibase 10.1007/JHEP12(2019)032} {\bibfield  {journal}
  {\bibinfo  {journal} {JHEP}\ }\textbf {\bibinfo {volume} {12}},\ \bibinfo
  {pages} {032} (\bibinfo {year} {2019})},\ \Eprint
  {http://arxiv.org/abs/1908.09845} {arXiv:1908.09845 [hep-ph]} \BibitemShut
  {NoStop}%
\bibitem [{\citenamefont {Bellazzini}\ \emph
  {et~al.}(2019{\natexlab{a}})\citenamefont {Bellazzini}, \citenamefont {Riva},
  \citenamefont {Serra},\ and\ \citenamefont {Sgarlata}}]{Bellazzini:2019bzh}%
  \BibitemOpen
  \bibfield  {author} {\bibinfo {author} {\bibfnamefont {B.}~\bibnamefont
  {Bellazzini}}, \bibinfo {author} {\bibfnamefont {F.}~\bibnamefont {Riva}},
  \bibinfo {author} {\bibfnamefont {J.}~\bibnamefont {Serra}}, \ and\ \bibinfo
  {author} {\bibfnamefont {F.}~\bibnamefont {Sgarlata}},\ }\href {\doibase
  10.1007/JHEP10(2019)189} {\bibfield  {journal} {\bibinfo  {journal} {JHEP}\
  }\textbf {\bibinfo {volume} {10}},\ \bibinfo {pages} {189} (\bibinfo {year}
  {2019}{\natexlab{a}})},\ \Eprint {http://arxiv.org/abs/1903.08664}
  {arXiv:1903.08664 [hep-th]} \BibitemShut {NoStop}%
\bibitem [{\citenamefont {Englert}\ \emph {et~al.}(2019)\citenamefont
  {Englert}, \citenamefont {Giudice}, \citenamefont {Greljo},\ and\
  \citenamefont {Mccullough}}]{Englert:2019zmt}%
  \BibitemOpen
  \bibfield  {author} {\bibinfo {author} {\bibfnamefont {C.}~\bibnamefont
  {Englert}}, \bibinfo {author} {\bibfnamefont {G.~F.}\ \bibnamefont
  {Giudice}}, \bibinfo {author} {\bibfnamefont {A.}~\bibnamefont {Greljo}}, \
  and\ \bibinfo {author} {\bibfnamefont {M.}~\bibnamefont {Mccullough}},\
  }\href {\doibase 10.1007/JHEP09(2019)041} {\bibfield  {journal} {\bibinfo
  {journal} {JHEP}\ }\textbf {\bibinfo {volume} {09}},\ \bibinfo {pages} {041}
  (\bibinfo {year} {2019})},\ \Eprint {http://arxiv.org/abs/1903.07725}
  {arXiv:1903.07725 [hep-ph]} \BibitemShut {NoStop}%
\bibitem [{\citenamefont {Trott}(2020)}]{Trott:2020ebl}%
  \BibitemOpen
  \bibfield  {author} {\bibinfo {author} {\bibfnamefont {T.}~\bibnamefont
  {Trott}},\ }\href@noop {} {\  (\bibinfo {year} {2020})},\ \Eprint
  {http://arxiv.org/abs/2011.10058} {arXiv:2011.10058 [hep-ph]} \BibitemShut
  {NoStop}%
\bibitem [{\citenamefont {Bonnefoy}\ \emph {et~al.}(2021)\citenamefont
  {Bonnefoy}, \citenamefont {Gendy},\ and\ \citenamefont
  {Grojean}}]{Bonnefoy:2020yee}%
  \BibitemOpen
  \bibfield  {author} {\bibinfo {author} {\bibfnamefont {Q.}~\bibnamefont
  {Bonnefoy}}, \bibinfo {author} {\bibfnamefont {E.}~\bibnamefont {Gendy}}, \
  and\ \bibinfo {author} {\bibfnamefont {C.}~\bibnamefont {Grojean}},\ }\href
  {\doibase 10.1007/JHEP04(2021)115} {\bibfield  {journal} {\bibinfo  {journal}
  {JHEP}\ }\textbf {\bibinfo {volume} {04}},\ \bibinfo {pages} {115} (\bibinfo
  {year} {2021})},\ \Eprint {http://arxiv.org/abs/2011.12855} {arXiv:2011.12855
  [hep-ph]} \BibitemShut {NoStop}%
\bibitem [{\citenamefont {Davighi}\ \emph {et~al.}(2021)\citenamefont
  {Davighi}, \citenamefont {Melville},\ and\ \citenamefont
  {You}}]{Davighi:2021osh}%
  \BibitemOpen
  \bibfield  {author} {\bibinfo {author} {\bibfnamefont {J.}~\bibnamefont
  {Davighi}}, \bibinfo {author} {\bibfnamefont {S.}~\bibnamefont {Melville}}, \
  and\ \bibinfo {author} {\bibfnamefont {T.}~\bibnamefont {You}},\ }\href@noop
  {} {\  (\bibinfo {year} {2021})},\ \Eprint {http://arxiv.org/abs/2108.06334}
  {arXiv:2108.06334 [hep-th]} \BibitemShut {NoStop}%
\bibitem [{\citenamefont {Chala}\ and\ \citenamefont
  {Santiago}(2021)}]{Chala:2021wpj}%
  \BibitemOpen
  \bibfield  {author} {\bibinfo {author} {\bibfnamefont {M.}~\bibnamefont
  {Chala}}\ and\ \bibinfo {author} {\bibfnamefont {J.}~\bibnamefont
  {Santiago}},\ }\href@noop {} {\  (\bibinfo {year} {2021})},\ \Eprint
  {http://arxiv.org/abs/2110.01624} {arXiv:2110.01624 [hep-ph]} \BibitemShut
  {NoStop}%
\bibitem [{\citenamefont {Azatov}\ \emph {et~al.}(2021)\citenamefont {Azatov},
  \citenamefont {Ghosh},\ and\ \citenamefont {Singh}}]{Azatov:2021ygj}%
  \BibitemOpen
  \bibfield  {author} {\bibinfo {author} {\bibfnamefont {A.}~\bibnamefont
  {Azatov}}, \bibinfo {author} {\bibfnamefont {D.}~\bibnamefont {Ghosh}}, \
  and\ \bibinfo {author} {\bibfnamefont {A.~H.}\ \bibnamefont {Singh}},\
  }\href@noop {} {\  (\bibinfo {year} {2021})},\ \Eprint
  {http://arxiv.org/abs/2112.02302} {arXiv:2112.02302 [hep-ph]} \BibitemShut
  {NoStop}%
\bibitem [{\citenamefont {Gruzinov}\ and\ \citenamefont
  {Kleban}(2007)}]{Gruzinov:2006ie}%
  \BibitemOpen
  \bibfield  {author} {\bibinfo {author} {\bibfnamefont {A.}~\bibnamefont
  {Gruzinov}}\ and\ \bibinfo {author} {\bibfnamefont {M.}~\bibnamefont
  {Kleban}},\ }\href {\doibase 10.1088/0264-9381/24/13/N02} {\bibfield
  {journal} {\bibinfo  {journal} {Class. Quant. Grav.}\ }\textbf {\bibinfo
  {volume} {24}},\ \bibinfo {pages} {3521} (\bibinfo {year} {2007})},\ \Eprint
  {http://arxiv.org/abs/hep-th/0612015} {arXiv:hep-th/0612015} \BibitemShut
  {NoStop}%
\bibitem [{\citenamefont {Bellazzini}\ \emph {et~al.}(2016)\citenamefont
  {Bellazzini}, \citenamefont {Cheung},\ and\ \citenamefont
  {Remmen}}]{Bellazzini:2015cra}%
  \BibitemOpen
  \bibfield  {author} {\bibinfo {author} {\bibfnamefont {B.}~\bibnamefont
  {Bellazzini}}, \bibinfo {author} {\bibfnamefont {C.}~\bibnamefont {Cheung}},
  \ and\ \bibinfo {author} {\bibfnamefont {G.~N.}\ \bibnamefont {Remmen}},\
  }\href {\doibase 10.1103/PhysRevD.93.064076} {\bibfield  {journal} {\bibinfo
  {journal} {Phys. Rev.}\ }\textbf {\bibinfo {volume} {D93}},\ \bibinfo {pages}
  {064076} (\bibinfo {year} {2016})},\ \Eprint
  {http://arxiv.org/abs/1509.00851} {arXiv:1509.00851 [hep-th]} \BibitemShut
  {NoStop}%
\bibitem [{\citenamefont {Bellazzini}\ \emph {et~al.}(2018)\citenamefont
  {Bellazzini}, \citenamefont {Riva}, \citenamefont {Serra},\ and\
  \citenamefont {Sgarlata}}]{Bellazzini:2017fep}%
  \BibitemOpen
  \bibfield  {author} {\bibinfo {author} {\bibfnamefont {B.}~\bibnamefont
  {Bellazzini}}, \bibinfo {author} {\bibfnamefont {F.}~\bibnamefont {Riva}},
  \bibinfo {author} {\bibfnamefont {J.}~\bibnamefont {Serra}}, \ and\ \bibinfo
  {author} {\bibfnamefont {F.}~\bibnamefont {Sgarlata}},\ }\href {\doibase
  10.1103/PhysRevLett.120.161101} {\bibfield  {journal} {\bibinfo  {journal}
  {Phys. Rev. Lett.}\ }\textbf {\bibinfo {volume} {120}},\ \bibinfo {pages}
  {161101} (\bibinfo {year} {2018})},\ \Eprint
  {http://arxiv.org/abs/1710.02539} {arXiv:1710.02539 [hep-th]} \BibitemShut
  {NoStop}%
\bibitem [{\citenamefont {Hamada}\ \emph {et~al.}(2019)\citenamefont {Hamada},
  \citenamefont {Noumi},\ and\ \citenamefont {Shiu}}]{Hamada:2018dde}%
  \BibitemOpen
  \bibfield  {author} {\bibinfo {author} {\bibfnamefont {Y.}~\bibnamefont
  {Hamada}}, \bibinfo {author} {\bibfnamefont {T.}~\bibnamefont {Noumi}}, \
  and\ \bibinfo {author} {\bibfnamefont {G.}~\bibnamefont {Shiu}},\ }\href
  {\doibase 10.1103/PhysRevLett.123.051601} {\bibfield  {journal} {\bibinfo
  {journal} {Phys. Rev. Lett.}\ }\textbf {\bibinfo {volume} {123}},\ \bibinfo
  {pages} {051601} (\bibinfo {year} {2019})},\ \Eprint
  {http://arxiv.org/abs/1810.03637} {arXiv:1810.03637 [hep-th]} \BibitemShut
  {NoStop}%
\bibitem [{\citenamefont {de~Rham}\ \emph
  {et~al.}(2018{\natexlab{b}})\citenamefont {de~Rham}, \citenamefont
  {Melville}, \citenamefont {Tolley},\ and\ \citenamefont
  {Zhou}}]{deRham:2018qqo}%
  \BibitemOpen
  \bibfield  {author} {\bibinfo {author} {\bibfnamefont {C.}~\bibnamefont
  {de~Rham}}, \bibinfo {author} {\bibfnamefont {S.}~\bibnamefont {Melville}},
  \bibinfo {author} {\bibfnamefont {A.~J.}\ \bibnamefont {Tolley}}, \ and\
  \bibinfo {author} {\bibfnamefont {S.-Y.}\ \bibnamefont {Zhou}},\ }\href@noop
  {} {\  (\bibinfo {year} {2018}{\natexlab{b}})},\ \Eprint
  {http://arxiv.org/abs/1804.10624} {arXiv:1804.10624 [hep-th]} \BibitemShut
  {NoStop}%
\bibitem [{\citenamefont {Alberte}\ \emph {et~al.}(2020)\citenamefont
  {Alberte}, \citenamefont {de~Rham}, \citenamefont {Momeni}, \citenamefont
  {Rumbutis},\ and\ \citenamefont {Tolley}}]{Alberte:2019xfh}%
  \BibitemOpen
  \bibfield  {author} {\bibinfo {author} {\bibfnamefont {L.}~\bibnamefont
  {Alberte}}, \bibinfo {author} {\bibfnamefont {C.}~\bibnamefont {de~Rham}},
  \bibinfo {author} {\bibfnamefont {A.}~\bibnamefont {Momeni}}, \bibinfo
  {author} {\bibfnamefont {J.}~\bibnamefont {Rumbutis}}, \ and\ \bibinfo
  {author} {\bibfnamefont {A.~J.}\ \bibnamefont {Tolley}},\ }\href {\doibase
  10.1007/JHEP03(2020)097} {\bibfield  {journal} {\bibinfo  {journal} {JHEP}\
  }\textbf {\bibinfo {volume} {03}},\ \bibinfo {pages} {097} (\bibinfo {year}
  {2020})},\ \Eprint {http://arxiv.org/abs/1910.11799} {arXiv:1910.11799
  [hep-th]} \BibitemShut {NoStop}%
\bibitem [{\citenamefont {Bellazzini}\ \emph
  {et~al.}(2019{\natexlab{b}})\citenamefont {Bellazzini}, \citenamefont
  {Lewandowski},\ and\ \citenamefont {Serra}}]{Bellazzini:2019xts}%
  \BibitemOpen
  \bibfield  {author} {\bibinfo {author} {\bibfnamefont {B.}~\bibnamefont
  {Bellazzini}}, \bibinfo {author} {\bibfnamefont {M.}~\bibnamefont
  {Lewandowski}}, \ and\ \bibinfo {author} {\bibfnamefont {J.}~\bibnamefont
  {Serra}},\ }\href {\doibase 10.1103/PhysRevLett.123.251103} {\bibfield
  {journal} {\bibinfo  {journal} {Phys. Rev. Lett.}\ }\textbf {\bibinfo
  {volume} {123}},\ \bibinfo {pages} {251103} (\bibinfo {year}
  {2019}{\natexlab{b}})},\ \Eprint {http://arxiv.org/abs/1902.03250}
  {arXiv:1902.03250 [hep-th]} \BibitemShut {NoStop}%
\bibitem [{\citenamefont {Kim}\ \emph {et~al.}(2019)\citenamefont {Kim},
  \citenamefont {Noumi}, \citenamefont {Takeuchi},\ and\ \citenamefont
  {Zhou}}]{Kim:2019wjo}%
  \BibitemOpen
  \bibfield  {author} {\bibinfo {author} {\bibfnamefont {S.}~\bibnamefont
  {Kim}}, \bibinfo {author} {\bibfnamefont {T.}~\bibnamefont {Noumi}}, \bibinfo
  {author} {\bibfnamefont {K.}~\bibnamefont {Takeuchi}}, \ and\ \bibinfo
  {author} {\bibfnamefont {S.}~\bibnamefont {Zhou}},\ }\href {\doibase
  10.1007/JHEP12(2019)107} {\bibfield  {journal} {\bibinfo  {journal} {JHEP}\
  }\textbf {\bibinfo {volume} {12}},\ \bibinfo {pages} {107} (\bibinfo {year}
  {2019})},\ \Eprint {http://arxiv.org/abs/1906.11840} {arXiv:1906.11840
  [hep-th]} \BibitemShut {NoStop}%
\bibitem [{\citenamefont {Tokuda}\ \emph {et~al.}(2020)\citenamefont {Tokuda},
  \citenamefont {Aoki},\ and\ \citenamefont {Hirano}}]{Tokuda:2020mlf}%
  \BibitemOpen
  \bibfield  {author} {\bibinfo {author} {\bibfnamefont {J.}~\bibnamefont
  {Tokuda}}, \bibinfo {author} {\bibfnamefont {K.}~\bibnamefont {Aoki}}, \ and\
  \bibinfo {author} {\bibfnamefont {S.}~\bibnamefont {Hirano}},\ }\href
  {\doibase 10.1007/JHEP11(2020)054} {\bibfield  {journal} {\bibinfo  {journal}
  {JHEP}\ }\textbf {\bibinfo {volume} {11}},\ \bibinfo {pages} {054} (\bibinfo
  {year} {2020})},\ \Eprint {http://arxiv.org/abs/2007.15009} {arXiv:2007.15009
  [hep-th]} \BibitemShut {NoStop}%
\bibitem [{\citenamefont {Herrero-Valea}\ \emph {et~al.}(2020)\citenamefont
  {Herrero-Valea}, \citenamefont {Santos-Garcia},\ and\ \citenamefont
  {Tokareva}}]{Herrero-Valea:2020wxz}%
  \BibitemOpen
  \bibfield  {author} {\bibinfo {author} {\bibfnamefont {M.}~\bibnamefont
  {Herrero-Valea}}, \bibinfo {author} {\bibfnamefont {R.}~\bibnamefont
  {Santos-Garcia}}, \ and\ \bibinfo {author} {\bibfnamefont {A.}~\bibnamefont
  {Tokareva}},\ }\href@noop {} {\  (\bibinfo {year} {2020})},\ \Eprint
  {http://arxiv.org/abs/2011.11652} {arXiv:2011.11652 [hep-th]} \BibitemShut
  {NoStop}%
\bibitem [{\citenamefont {Bern}\ \emph {et~al.}(2021)\citenamefont {Bern},
  \citenamefont {Kosmopoulos},\ and\ \citenamefont {Zhiboedov}}]{Bern:2021ppb}%
  \BibitemOpen
  \bibfield  {author} {\bibinfo {author} {\bibfnamefont {Z.}~\bibnamefont
  {Bern}}, \bibinfo {author} {\bibfnamefont {D.}~\bibnamefont {Kosmopoulos}}, \
  and\ \bibinfo {author} {\bibfnamefont {A.}~\bibnamefont {Zhiboedov}},\
  }\href@noop {} {\  (\bibinfo {year} {2021})},\ \Eprint
  {http://arxiv.org/abs/2103.12728} {arXiv:2103.12728 [hep-th]} \BibitemShut
  {NoStop}%
\bibitem [{\citenamefont {Camanho}\ \emph {et~al.}(2016)\citenamefont
  {Camanho}, \citenamefont {Edelstein}, \citenamefont {Maldacena},\ and\
  \citenamefont {Zhiboedov}}]{Camanho:2014apa}%
  \BibitemOpen
  \bibfield  {author} {\bibinfo {author} {\bibfnamefont {X.~O.}\ \bibnamefont
  {Camanho}}, \bibinfo {author} {\bibfnamefont {J.~D.}\ \bibnamefont
  {Edelstein}}, \bibinfo {author} {\bibfnamefont {J.}~\bibnamefont
  {Maldacena}}, \ and\ \bibinfo {author} {\bibfnamefont {A.}~\bibnamefont
  {Zhiboedov}},\ }\href {\doibase 10.1007/JHEP02(2016)020} {\bibfield
  {journal} {\bibinfo  {journal} {JHEP}\ }\textbf {\bibinfo {volume} {02}},\
  \bibinfo {pages} {020} (\bibinfo {year} {2016})},\ \Eprint
  {http://arxiv.org/abs/1407.5597} {arXiv:1407.5597 [hep-th]} \BibitemShut
  {NoStop}%
\bibitem [{\citenamefont {Afkhami-Jeddi}\ \emph {et~al.}(2018)\citenamefont
  {Afkhami-Jeddi}, \citenamefont {Kundu},\ and\ \citenamefont
  {Tajdini}}]{Afkhami-Jeddi:2018apj}%
  \BibitemOpen
  \bibfield  {author} {\bibinfo {author} {\bibfnamefont {N.}~\bibnamefont
  {Afkhami-Jeddi}}, \bibinfo {author} {\bibfnamefont {S.}~\bibnamefont
  {Kundu}}, \ and\ \bibinfo {author} {\bibfnamefont {A.}~\bibnamefont
  {Tajdini}},\ }\href@noop {} {\  (\bibinfo {year} {2018})},\ \Eprint
  {http://arxiv.org/abs/1811.01952} {arXiv:1811.01952 [hep-th]} \BibitemShut
  {NoStop}%
\bibitem [{\citenamefont {Bellazzini}\ \emph {et~al.}(2021)\citenamefont
  {Bellazzini}, \citenamefont {Isabella}, \citenamefont {Lewandowski},\ and\
  \citenamefont {Sgarlata}}]{Bellazzini:2021shn}%
  \BibitemOpen
  \bibfield  {author} {\bibinfo {author} {\bibfnamefont {B.}~\bibnamefont
  {Bellazzini}}, \bibinfo {author} {\bibfnamefont {G.}~\bibnamefont
  {Isabella}}, \bibinfo {author} {\bibfnamefont {M.}~\bibnamefont
  {Lewandowski}}, \ and\ \bibinfo {author} {\bibfnamefont {F.}~\bibnamefont
  {Sgarlata}},\ }\href@noop {} {\  (\bibinfo {year} {2021})},\ \Eprint
  {http://arxiv.org/abs/2108.05896} {arXiv:2108.05896 [hep-th]} \BibitemShut
  {NoStop}%
\bibitem [{\citenamefont {Arkani-Hamed}\ \emph {et~al.}(2020)\citenamefont
  {Arkani-Hamed}, \citenamefont {Huang},\ and\ \citenamefont
  {Huang}}]{Arkani-Hamed:2020blm}%
  \BibitemOpen
  \bibfield  {author} {\bibinfo {author} {\bibfnamefont {N.}~\bibnamefont
  {Arkani-Hamed}}, \bibinfo {author} {\bibfnamefont {T.-C.}\ \bibnamefont
  {Huang}}, \ and\ \bibinfo {author} {\bibfnamefont {Y.-T.}\ \bibnamefont
  {Huang}},\ }\href@noop {} {\  (\bibinfo {year} {2020})},\ \Eprint
  {http://arxiv.org/abs/2012.15849} {arXiv:2012.15849 [hep-th]} \BibitemShut
  {NoStop}%
\bibitem [{\citenamefont {Bellazzini}\ \emph {et~al.}(2020)\citenamefont
  {Bellazzini}, \citenamefont {Elias~Mir\'o}, \citenamefont {Rattazzi},
  \citenamefont {Riembau},\ and\ \citenamefont {Riva}}]{Bellazzini:2020cot}%
  \BibitemOpen
  \bibfield  {author} {\bibinfo {author} {\bibfnamefont {B.}~\bibnamefont
  {Bellazzini}}, \bibinfo {author} {\bibfnamefont {J.}~\bibnamefont
  {Elias~Mir\'o}}, \bibinfo {author} {\bibfnamefont {R.}~\bibnamefont
  {Rattazzi}}, \bibinfo {author} {\bibfnamefont {M.}~\bibnamefont {Riembau}}, \
  and\ \bibinfo {author} {\bibfnamefont {F.}~\bibnamefont {Riva}},\ }\href@noop
  {} {\  (\bibinfo {year} {2020})},\ \Eprint {http://arxiv.org/abs/2011.00037}
  {arXiv:2011.00037 [hep-th]} \BibitemShut {NoStop}%
\bibitem [{\citenamefont {Tolley}\ \emph {et~al.}(2020)\citenamefont {Tolley},
  \citenamefont {Wang},\ and\ \citenamefont {Zhou}}]{Tolley:2020gtv}%
  \BibitemOpen
  \bibfield  {author} {\bibinfo {author} {\bibfnamefont {A.~J.}\ \bibnamefont
  {Tolley}}, \bibinfo {author} {\bibfnamefont {Z.-Y.}\ \bibnamefont {Wang}}, \
  and\ \bibinfo {author} {\bibfnamefont {S.-Y.}\ \bibnamefont {Zhou}},\
  }\href@noop {} {\  (\bibinfo {year} {2020})},\ \Eprint
  {http://arxiv.org/abs/2011.02400} {arXiv:2011.02400 [hep-th]} \BibitemShut
  {NoStop}%
\bibitem [{\citenamefont {Caron-Huot}\ and\ \citenamefont
  {Van~Duong}(2021)}]{Caron-Huot:2020cmc}%
  \BibitemOpen
  \bibfield  {author} {\bibinfo {author} {\bibfnamefont {S.}~\bibnamefont
  {Caron-Huot}}\ and\ \bibinfo {author} {\bibfnamefont {V.}~\bibnamefont
  {Van~Duong}},\ }\href {\doibase 10.1007/JHEP05(2021)280} {\bibfield
  {journal} {\bibinfo  {journal} {JHEP}\ }\textbf {\bibinfo {volume} {05}},\
  \bibinfo {pages} {280} (\bibinfo {year} {2021})},\ \Eprint
  {http://arxiv.org/abs/2011.02957} {arXiv:2011.02957 [hep-th]} \BibitemShut
  {NoStop}%
\bibitem [{\citenamefont {Sinha}\ and\ \citenamefont
  {Zahed}(2021)}]{Sinha:2020win}%
  \BibitemOpen
  \bibfield  {author} {\bibinfo {author} {\bibfnamefont {A.}~\bibnamefont
  {Sinha}}\ and\ \bibinfo {author} {\bibfnamefont {A.}~\bibnamefont {Zahed}},\
  }\href {\doibase 10.1103/PhysRevLett.126.181601} {\bibfield  {journal}
  {\bibinfo  {journal} {Phys. Rev. Lett.}\ }\textbf {\bibinfo {volume} {126}},\
  \bibinfo {pages} {181601} (\bibinfo {year} {2021})},\ \Eprint
  {http://arxiv.org/abs/2012.04877} {arXiv:2012.04877 [hep-th]} \BibitemShut
  {NoStop}%
\bibitem [{\citenamefont {Caron-Huot}\ \emph {et~al.}(2021)\citenamefont
  {Caron-Huot}, \citenamefont {Mazac}, \citenamefont {Rastelli},\ and\
  \citenamefont {Simmons-Duffin}}]{Caron-Huot:2021rmr}%
  \BibitemOpen
  \bibfield  {author} {\bibinfo {author} {\bibfnamefont {S.}~\bibnamefont
  {Caron-Huot}}, \bibinfo {author} {\bibfnamefont {D.}~\bibnamefont {Mazac}},
  \bibinfo {author} {\bibfnamefont {L.}~\bibnamefont {Rastelli}}, \ and\
  \bibinfo {author} {\bibfnamefont {D.}~\bibnamefont {Simmons-Duffin}},\
  }\href@noop {} {\  (\bibinfo {year} {2021})},\ \Eprint
  {http://arxiv.org/abs/2102.08951} {arXiv:2102.08951 [hep-th]} \BibitemShut
  {NoStop}%
\bibitem [{\citenamefont {Chiang}\ \emph {et~al.}(2021)\citenamefont {Chiang},
  \citenamefont {Huang}, \citenamefont {Li}, \citenamefont {Rodina},\ and\
  \citenamefont {Weng}}]{Chiang:2021ziz}%
  \BibitemOpen
  \bibfield  {author} {\bibinfo {author} {\bibfnamefont {L.-Y.}\ \bibnamefont
  {Chiang}}, \bibinfo {author} {\bibfnamefont {Y.-t.}\ \bibnamefont {Huang}},
  \bibinfo {author} {\bibfnamefont {W.}~\bibnamefont {Li}}, \bibinfo {author}
  {\bibfnamefont {L.}~\bibnamefont {Rodina}}, \ and\ \bibinfo {author}
  {\bibfnamefont {H.-C.}\ \bibnamefont {Weng}},\ }\href@noop {} {\  (\bibinfo
  {year} {2021})},\ \Eprint {http://arxiv.org/abs/2105.02862} {arXiv:2105.02862
  [hep-th]} \BibitemShut {NoStop}%
\bibitem [{\citenamefont {Henriksson}\ \emph {et~al.}(2021)\citenamefont
  {Henriksson}, \citenamefont {McPeak}, \citenamefont {Russo},\ and\
  \citenamefont {Vichi}}]{Henriksson:2021ymi}%
  \BibitemOpen
  \bibfield  {author} {\bibinfo {author} {\bibfnamefont {J.}~\bibnamefont
  {Henriksson}}, \bibinfo {author} {\bibfnamefont {B.}~\bibnamefont {McPeak}},
  \bibinfo {author} {\bibfnamefont {F.}~\bibnamefont {Russo}}, \ and\ \bibinfo
  {author} {\bibfnamefont {A.}~\bibnamefont {Vichi}},\ }\href@noop {} {\
  (\bibinfo {year} {2021})},\ \Eprint {http://arxiv.org/abs/2107.13009}
  {arXiv:2107.13009 [hep-th]} \BibitemShut {NoStop}%
\bibitem [{\citenamefont {Nicolis}\ \emph {et~al.}(2009)\citenamefont
  {Nicolis}, \citenamefont {Rattazzi},\ and\ \citenamefont
  {Trincherini}}]{Nicolis:2008in}%
  \BibitemOpen
  \bibfield  {author} {\bibinfo {author} {\bibfnamefont {A.}~\bibnamefont
  {Nicolis}}, \bibinfo {author} {\bibfnamefont {R.}~\bibnamefont {Rattazzi}}, \
  and\ \bibinfo {author} {\bibfnamefont {E.}~\bibnamefont {Trincherini}},\
  }\href {\doibase 10.1103/PhysRevD.79.064036} {\bibfield  {journal} {\bibinfo
  {journal} {Phys. Rev.}\ }\textbf {\bibinfo {volume} {D79}},\ \bibinfo {pages}
  {064036} (\bibinfo {year} {2009})},\ \Eprint {http://arxiv.org/abs/0811.2197}
  {arXiv:0811.2197 [hep-th]} \BibitemShut {NoStop}%
\bibitem [{\citenamefont {Froissart}(1961)}]{Froissart:1961ux}%
  \BibitemOpen
  \bibfield  {author} {\bibinfo {author} {\bibfnamefont {M.}~\bibnamefont
  {Froissart}},\ }\href {\doibase 10.1103/PhysRev.123.1053} {\bibfield
  {journal} {\bibinfo  {journal} {Phys. Rev.}\ }\textbf {\bibinfo {volume}
  {123}},\ \bibinfo {pages} {1053} (\bibinfo {year} {1961})}\BibitemShut
  {NoStop}%
\bibitem [{\citenamefont {Martin}(1963)}]{Martin:1962rt}%
  \BibitemOpen
  \bibfield  {author} {\bibinfo {author} {\bibfnamefont {A.}~\bibnamefont
  {Martin}},\ }\href {\doibase 10.1103/PhysRev.129.1432} {\bibfield  {journal}
  {\bibinfo  {journal} {Phys. Rev.}\ }\textbf {\bibinfo {volume} {129}},\
  \bibinfo {pages} {1432} (\bibinfo {year} {1963})}\BibitemShut {NoStop}%
\bibitem [{\citenamefont {Jin}\ and\ \citenamefont
  {Martin}(1964)}]{Jin:1964zza}%
  \BibitemOpen
  \bibfield  {author} {\bibinfo {author} {\bibfnamefont {Y.~S.}\ \bibnamefont
  {Jin}}\ and\ \bibinfo {author} {\bibfnamefont {A.}~\bibnamefont {Martin}},\
  }\href {\doibase 10.1103/PhysRev.135.B1375} {\bibfield  {journal} {\bibinfo
  {journal} {Phys. Rev.}\ }\textbf {\bibinfo {volume} {135}},\ \bibinfo {pages}
  {B1375} (\bibinfo {year} {1964})}\BibitemShut {NoStop}%
\bibitem [{\citenamefont {Martin}(1965)}]{Martin:1965jj}%
  \BibitemOpen
  \bibfield  {author} {\bibinfo {author} {\bibfnamefont {A.}~\bibnamefont
  {Martin}},\ }\href {\doibase 10.1007/BF02720568} {\bibfield  {journal}
  {\bibinfo  {journal} {Nuovo Cim.}\ }\textbf {\bibinfo {volume} {A42}},\
  \bibinfo {pages} {930} (\bibinfo {year} {1965})}\BibitemShut {NoStop}%
\bibitem [{\citenamefont {Martin}(1969)}]{Martin:1969ina}%
  \BibitemOpen
  \bibfield  {author} {\bibinfo {author} {\bibfnamefont {A.}~\bibnamefont
  {Martin}},\ }\href {\doibase 10.1007/BFb0101043} {\emph {\bibinfo {title}
  {{Scattering Theory: Unitarity, Analyticity and Crossing}}}},\ Vol.~\bibinfo
  {volume} {3}\ (\bibinfo {year} {1969})\BibitemShut {NoStop}%
\bibitem [{\citenamefont {Sommer}(1970)}]{Sommer:1970mr}%
  \BibitemOpen
  \bibfield  {author} {\bibinfo {author} {\bibfnamefont {G.}~\bibnamefont
  {Sommer}},\ }\href {\doibase 10.1002/prop.19700181102} {\bibfield  {journal}
  {\bibinfo  {journal} {Fortsch. Phys.}\ }\textbf {\bibinfo {volume} {18}},\
  \bibinfo {pages} {577} (\bibinfo {year} {1970})}\BibitemShut {NoStop}%
\bibitem [{\citenamefont {Lasserre}(2009)}]{lasserreJB}%
  \BibitemOpen
  \bibfield  {author} {\bibinfo {author} {\bibfnamefont {J.}~\bibnamefont
  {Lasserre}},\ }\href@noop {} {\ \bibinfo {series} {Series on Optimization and
  Its Applications: Volume 1} (\bibinfo {year} {2009})}\BibitemShut {NoStop}%
\bibitem [{\citenamefont {Schmuedgen}(1991)}]{schmudgen91}%
  \BibitemOpen
  \bibfield  {author} {\bibinfo {author} {\bibfnamefont {K.}~\bibnamefont
  {Schmuedgen}},\ }\href@noop {} {\bibfield  {journal} {\bibinfo  {journal}
  {Math. Ann.}\ }\textbf {\bibinfo {volume} {289}} (\bibinfo {year}
  {1991})}\BibitemShut {NoStop}%
\bibitem [{\citenamefont {Guerrieri}\ and\ \citenamefont
  {Sever}(2021)}]{Guerrieri:2021tak}%
  \BibitemOpen
  \bibfield  {author} {\bibinfo {author} {\bibfnamefont {A.}~\bibnamefont
  {Guerrieri}}\ and\ \bibinfo {author} {\bibfnamefont {A.}~\bibnamefont
  {Sever}},\ }\href@noop {} {\  (\bibinfo {year} {2021})},\ \Eprint
  {http://arxiv.org/abs/2106.10257} {arXiv:2106.10257 [hep-th]} \BibitemShut
  {NoStop}%
\bibitem [{\citenamefont {Paulos}\ \emph
  {et~al.}(2017{\natexlab{a}})\citenamefont {Paulos}, \citenamefont
  {Penedones}, \citenamefont {Toledo}, \citenamefont {van Rees},\ and\
  \citenamefont {Vieira}}]{Paulos:2016fap}%
  \BibitemOpen
  \bibfield  {author} {\bibinfo {author} {\bibfnamefont {M.~F.}\ \bibnamefont
  {Paulos}}, \bibinfo {author} {\bibfnamefont {J.}~\bibnamefont {Penedones}},
  \bibinfo {author} {\bibfnamefont {J.}~\bibnamefont {Toledo}}, \bibinfo
  {author} {\bibfnamefont {B.~C.}\ \bibnamefont {van Rees}}, \ and\ \bibinfo
  {author} {\bibfnamefont {P.}~\bibnamefont {Vieira}},\ }\href {\doibase
  10.1007/JHEP11(2017)133} {\bibfield  {journal} {\bibinfo  {journal} {JHEP}\
  }\textbf {\bibinfo {volume} {11}},\ \bibinfo {pages} {133} (\bibinfo {year}
  {2017}{\natexlab{a}})},\ \Eprint {http://arxiv.org/abs/1607.06109}
  {arXiv:1607.06109 [hep-th]} \BibitemShut {NoStop}%
\bibitem [{\citenamefont {Paulos}\ \emph
  {et~al.}(2017{\natexlab{b}})\citenamefont {Paulos}, \citenamefont
  {Penedones}, \citenamefont {Toledo}, \citenamefont {van Rees},\ and\
  \citenamefont {Vieira}}]{Paulos:2016but}%
  \BibitemOpen
  \bibfield  {author} {\bibinfo {author} {\bibfnamefont {M.~F.}\ \bibnamefont
  {Paulos}}, \bibinfo {author} {\bibfnamefont {J.}~\bibnamefont {Penedones}},
  \bibinfo {author} {\bibfnamefont {J.}~\bibnamefont {Toledo}}, \bibinfo
  {author} {\bibfnamefont {B.~C.}\ \bibnamefont {van Rees}}, \ and\ \bibinfo
  {author} {\bibfnamefont {P.}~\bibnamefont {Vieira}},\ }\href {\doibase
  10.1007/JHEP11(2017)143} {\bibfield  {journal} {\bibinfo  {journal} {JHEP}\
  }\textbf {\bibinfo {volume} {11}},\ \bibinfo {pages} {143} (\bibinfo {year}
  {2017}{\natexlab{b}})},\ \Eprint {http://arxiv.org/abs/1607.06110}
  {arXiv:1607.06110 [hep-th]} \BibitemShut {NoStop}%
\bibitem [{\citenamefont {Paulos}\ \emph
  {et~al.}(2017{\natexlab{c}})\citenamefont {Paulos}, \citenamefont
  {Penedones}, \citenamefont {Toledo}, \citenamefont {van Rees},\ and\
  \citenamefont {Vieira}}]{Paulos:2017fhb}%
  \BibitemOpen
  \bibfield  {author} {\bibinfo {author} {\bibfnamefont {M.~F.}\ \bibnamefont
  {Paulos}}, \bibinfo {author} {\bibfnamefont {J.}~\bibnamefont {Penedones}},
  \bibinfo {author} {\bibfnamefont {J.}~\bibnamefont {Toledo}}, \bibinfo
  {author} {\bibfnamefont {B.~C.}\ \bibnamefont {van Rees}}, \ and\ \bibinfo
  {author} {\bibfnamefont {P.}~\bibnamefont {Vieira}},\ }\href@noop {} {\
  (\bibinfo {year} {2017}{\natexlab{c}})},\ \Eprint
  {http://arxiv.org/abs/1708.06765} {arXiv:1708.06765 [hep-th]} \BibitemShut
  {NoStop}%
\bibitem [{\citenamefont {Bros}\ \emph {et~al.}(1965)\citenamefont {Bros},
  \citenamefont {Epstein},\ and\ \citenamefont {Glaser}}]{Bros:1965kbd}%
  \BibitemOpen
  \bibfield  {author} {\bibinfo {author} {\bibfnamefont {J.}~\bibnamefont
  {Bros}}, \bibinfo {author} {\bibfnamefont {H.}~\bibnamefont {Epstein}}, \
  and\ \bibinfo {author} {\bibfnamefont {V.}~\bibnamefont {Glaser}},\ }\href
  {\doibase 10.1007/BF01646307} {\bibfield  {journal} {\bibinfo  {journal}
  {Commun. Math. Phys.}\ }\textbf {\bibinfo {volume} {1}},\ \bibinfo {pages}
  {240} (\bibinfo {year} {1965})}\BibitemShut {NoStop}%
\end{thebibliography}%

\end{document}